\documentclass[twocolumn,superscriptaddress,prb,preprintnumbers,nobibnotes,aps]{revtex4-2}  %,showpacs
\usepackage{graphicx}% Include figure files
\usepackage[caption=false,font=large]{subfig}
\usepackage{floatrow}
\usepackage{epstopdf}
\usepackage{dcolumn}% Align table columns on decimal point
\usepackage{multirow}
\usepackage{hyperref}% add hypertext capabilities; optional: [hidelinks]
\hypersetup{
  colorlinks  = true, %Colours links instead of boxes
  urlcolor   = blue, %Colour for external hyperlinks
  linkcolor  = blue, %Colour of internal links
  citecolor  = blue %Colour of citations
}
\usepackage{xcolor}
\usepackage{soul}
\usepackage{natbib}
\usepackage{amsmath}
\usepackage{rotating}
\setcitestyle{square,numbers}

\newcommand{\gc}{$^\circ$C}

\begin{document}

\title{Crystal Growth and Anisotropic Magnetic Properties of Quasi-2D (Fe$_{1-x}$Ni$_{x}$)$_{2}$P$_{2}$S$_{6}$}	

\author{S.~Selter}
\thanks{These authors contributed equally to this work.}
\affiliation{Institute for Solid State Research, Leibniz IFW Dresden, Helmholtzstr. 20, 01069 Dresden, Germany}
\affiliation{Institute of Solid State and Materials Physics, Technische Universit\"at Dresden, 01062 Dresden, Germany}
\author{Y.~Shemerliuk}
\thanks{These authors contributed equally to this work.}
\affiliation{Institute for Solid State Research, Leibniz IFW Dresden, Helmholtzstr. 20, 01069 Dresden, Germany}
\author{M.-I.~Sturza}
\affiliation{Institute for Solid State Research, Leibniz IFW Dresden, Helmholtzstr. 20, 01069 Dresden, Germany}
\author{A.~U.~B.~Wolter}
\affiliation{Institute for Solid State Research, Leibniz IFW Dresden, Helmholtzstr. 20, 01069 Dresden, Germany}
\author{B.~B\"{u}chner}
\affiliation{Institute for Solid State Research, Leibniz IFW Dresden, Helmholtzstr. 20, 01069 Dresden, Germany}
\affiliation{Institute of Solid State and Materials Physics and W\"urzburg-Dresden Cluster of Excellence
ct.qmat, Technische Universit\"at Dresden, 01062 Dresden, Germany}
\author{S.~Aswartham}
\email{s.aswartham@ifw-dresden.de}
\affiliation{Institute for Solid State Research, Leibniz IFW Dresden, Helmholtzstr. 20, 01069 Dresden, Germany}
\date{\today}

\begin{abstract}
We report the crystal growth by chemical vapor transport together with a thorough structural and magnetic characterization of the quasi-2D magnets (Fe$_{1-x}$Ni$_x$)$_2$P$_2$S$_6$ with $x_\textrm{Ni} = \textrm{0, 0.3, 0.5, 0.7, 0.9 \& 1}$. As-grown crystals exhibit a layered morphology with weak van der Waals interactions between layers parallel to the crystallographic $ab$ plane of the monoclinic structure in the space group $C2/m$ (No.~12). Magnetization measurements reveal an antiferromagnetic ground state for all grown crystals. In the ordered state, the magnetization along different crystallographic directions is in agreement with an Ising anisotropy for 0 $\leq$ $x$ $\leq$ 0.9 and only for Ni$_2$P$_2$S$_6$ a different behavior is observed which is line with an anisotropic Heisenberg or XXZ model description. This striking abrupt change of anisotropy at a critical Ni concentration is coupled to a change in the width of the maximum in the magnetization around $T_\textrm{max}$. Furthermore, all intermediate compounds 0 $<$ $x$ $\leq$ 0.9 exhibit an anisotropic magnetization already in the paramagnetic state similar to the parent compound Fe$_2$P$_2$S$_6$.
\end{abstract}


\maketitle

\section{Introduction}

Since the experimental verification of stable ferromagnetic order in monolayer and few-layer samples of CrI$_3$~\cite{Huang2017} and Cr$_2$Ge$_2$Te$_6$~\cite{CGong2017}, magnetic two-dimensional (2D) systems received rapidly growing attention~\cite{NatNano2018}. Such low-dimensional magnetic systems are promising for the use in spintronics, miniaturization of functional devices as well as to supplement the well-studied non-magnetic 2D materials in heterostructures via induced magnetic interactions and phenomena~\cite{Wang2018,Gibertini2019,Zhong2017,Song2019,Wang2018a,Samarth2017}. For an optimized design of such magnetic 2D materials, a reliable understanding of magnetic order in low-dimensional systems and its interplay with magnetic anisotropy in the context of the Mermin-Wagner theorem are necessary~\cite{Mermin1966}. Magnetic 2D materials may also provide a platform to study exotic short-range correlation-driven effects like the Berezinsky-Kosterlitz-Thouless transition predicted for the XY spin model in 2D systems~\cite{Kosterlitz1973,Berezinsky1970}.

To study the fundamental magnetic properties of 2D systems, the family of $M_2$P$_2$S$_6$ compounds ($M = \textrm{V, Mn, Fe, Co, Ni}$) offers a series of isostructural van der Waals layered compounds, all exhibiting antiferromagnetic order as bulk material but with different magnetic anisotropies~\cite{Brec1986}. For example, Mn$_2$P$_2$S$_6$ is a Heisenberg antiferromagnet below a N\'{e}el temperature $T_\textrm{N} = 78$\,K~\cite{PJoy1992,Wildes1998}, Fe$_2$P$_2$S$_6$ is an Ising antiferromagnet below $T_\textrm{N} = 123$\,K~\cite{PJoy1992,DLancon2016} and Ni$_2$P$_2$S$_6$ exhibits an antiferromagnetic state below $T_\textrm{N} = 155$\,K which can be described by an anisotropic Heisenberg model or the XXZ model~\cite{PJoy1992,Lancon2018,Kim2019}. For monolayer samples of Ni$_2$P$_2$S$_6$, magnetic order breaks down~\cite{Kim2019}, while it survives in the strongly anisotropic atomically thin Fe$_2$P$_2$S$_6$~\cite{Lee2016}, following the expectations from the Mermin-Wagner theorem.

Tuning the magnetic properties with chemical substitution, i.e., gradually substituting Mn with Fe in Mn$_2$P$_2$S$_6$ or Ni with Fe in Ni$_2$P$_2$S$_6$ may allow to tune the magnetic anisotropy in the system. This will allow to investigate the interplay between magnetic anisotropy, magnetic interactions and spin fluctuations in low-dimensional systems. Along this line, it may be furthermore possible to tune the system to the sweet spot of a critical magnetic anisotropy strength at which the low-dimensional magnetic system is dominated by short-range correlations, as e.g. described by Berezinsky and Kosterlitz~\cite{Kosterlitz1973,Berezinsky1970}.

Indeed, the work on the substitution series (Mn$_{1-x}$Fe$_{x}$)$_2$P$_2$S$_6$ by Masubuchi~\textit{et~al.}~\cite{Masubuchi2008} demonstrated the fundamental feasibility of substituting between two magnetic $M_2$P$_2$S$_6$ compounds including a solid solution behavior, which is essential for a gradual evolution of the physical properties as function of the degree of substitution. However, the evolution of magnetic properties in this specific substitution series exhibits a breakdown of long-range magnetic order and the existence of a spin-glass state around (Mn$_{0.5}$Fe$_{0.5}$)$_2$P$_2$S$_6$ rather than a gradual evolution between the magnetic characteristics of both parent compounds.

In contrast to (Mn$_{1-x}$Fe$_{x}$)$_2$P$_2$S$_6$, an initial work on polycrystalline samples of (Fe$_{1-x}$Ni$_x$)$_2$P$_2$S$_6$ \cite{RRao1992} proved the existence of long-range magnetic order for the full series, making this substitution series a promising candidate for deliberately tuning the magnetic anisotropy. However, to reliably investigate the magnetic anisotropy of a material, single crystalline samples are necessary allowing magnetic measurements with magnetic fields applied along different crystallographic directions.

In this work we report the crystal growth of the quasi-2D van der Waals layered series (Fe$_{1-x}$Ni$_x$)$_2$P$_2$S$_6$. Both parent compounds as well as several intermediate compounds ($x_\textrm{Ni} =$ 0.3, 0.5, 0.7 {\&} 0.9) were grown using the chemical vapor transport (CVT) technique with iodine as transport agent. Crystals of all degrees of Ni-substitution were thoroughly characterized by scanning electron microscopy and energy dispersive X-ray spectroscopy regarding morphology, chemical homogeneity and elemental composition as well as by single crystal X-ray diffraction and powder X-ray diffraction regarding the crystal structure and lattice parameters. An investigation of the bulk magnetic properties is presented based on magnetization studies performed along all principle crystallographic axis, illustrating the effect of Fe/Ni substitution on the magnetic anisotropy in (Fe$_{1-x}$Ni$_x$)$_2$P$_2$S$_6$ single crystals.

\section{Crystal Growth and Methods}

Single crystals were obtained by a two-step process: First, stoichiometric polycrystalline powders were obtained by solid state synthesis from the elements. In a second step, crystals were grown from these polycrystalline precursors by CVT. This two-step process has proven necessary to obtain homogeneous single crystals with the desired degree of Ni-substitution. A similar approach was used by us for the crystal growth in the closely related substitution series (Mn$_{1-x}$Ni$_{x}$)$_2$P$_2$S$_6$ \cite{Shemerliuk2021}.

\subsection{Synthesis of the polycrystalline precursors}

The elemental constituents iron (powder -70 mesh, Acros Organics, 99\%), nickel (powder -100 mesh, Sigma Aldrich, 99.99\%), red phosphorus (lumps, Alfa Aesar, 99.999\%) and sulfur (pieces, Alfa Aesar, 99.999\%) were weighed out in stoichiometric quantities with respect to (Fe$_{1-x}$Ni$_x$)$_2$P$_2$S$_6$ ($x_\textrm{Ni}=$ 0,~0.3,~0.5,~0.7,~0.9{\&}~1). The elemental mixtures were homogenized in an agate mortar and pressed to pellets (1\,cm diameter) at approximately 25\,kN using a hydraulic press. Typically 2\,g of pellets were loaded in a quartz ampule (10\,mm inner diameter, 3\,mm wall thickness). All preparation steps up to here were performed under argon atmosphere inside a glove box. Quartz ampules were thoroughly cleaned by washing with destilled water, rinsing with ethanol or isopropanol and, subsequently, baked out at 800\,\gc\ for at least 12\,h immediately prior to use to avoid contamination by (adsorbed) water. The ampule was then sealed under an internal pressure of approximately 0.3\,bar Ar (at 20\,\gc). Finally, the pellets were heat-treated in a tube furnace. Initially, the furnace was heated to 300\,\gc\ with 50\,\gc/h and dwelled for 24\,h to ensure pre-reaction of the volatile elements P and S with the transition elements. Then it was heated to 600\,\gc\ with 100\,\gc/h and dwelled for 72\,h. After this, the furnace was turned off and the pellets were furnace-cooled to room temperature. The pellets were pulverized under argon atmosphere and the formation of the monoclinic phase was confirmed by powder X-ray diffraction (pXRD) measurements.

\subsection{Crystal growth}

\subsubsection{Fe$_2$P$_2$S$_6$}

To grow single crystals, 0.5\,g of pre-reacted starting material of Fe$_2$P$_2$S$_6$ was loaded in a quartz ampule (6\,mm inner diameter, 2\,mm wall thickness, cleaned as described above) as a fine powder together with a small amount (approx. 5\,mol-\% with respect to the transition metals) of the transport agent iodine (resublimed crystals, Alfa Aesar, 99.9985\%). This was done in a glove box under argon atmosphere. The ampule containing the starting materials was then transferred to a vacuum pump and evacuated to a residual pressure of $10^{-8}$\,bar. To suppress the unintended sublimation of the transport agent during evacuation, the material containing end of the ampule was cooled by attaching a small Dewar flask filled with liquid nitrogen. After stabilizing the desired internal pressure, the valve to the vacuum pump was closed, the cooling was stopped and the ampule was sealed under static pressure at a length of approximately 12\,cm by a oxyhydrogen flame. The ampule was placed horizontally in a two-zone tube furnace in such a way that the elemental mixture was only at one side of the ampoule which is called the charge side. The same temperature profile as reported by Rule~\textit{et~al.}~\cite{Rule2002} was applied. However, the obtained crystals were too small for a detailed investigation of the anisotropic magnetic properties. Thus, we optimized the growth conditions as reported hereafter and were able to grow single crystals with a mass of up to 5\,mg.

\begin{figure}[htb]
\includegraphics[width=\columnwidth]{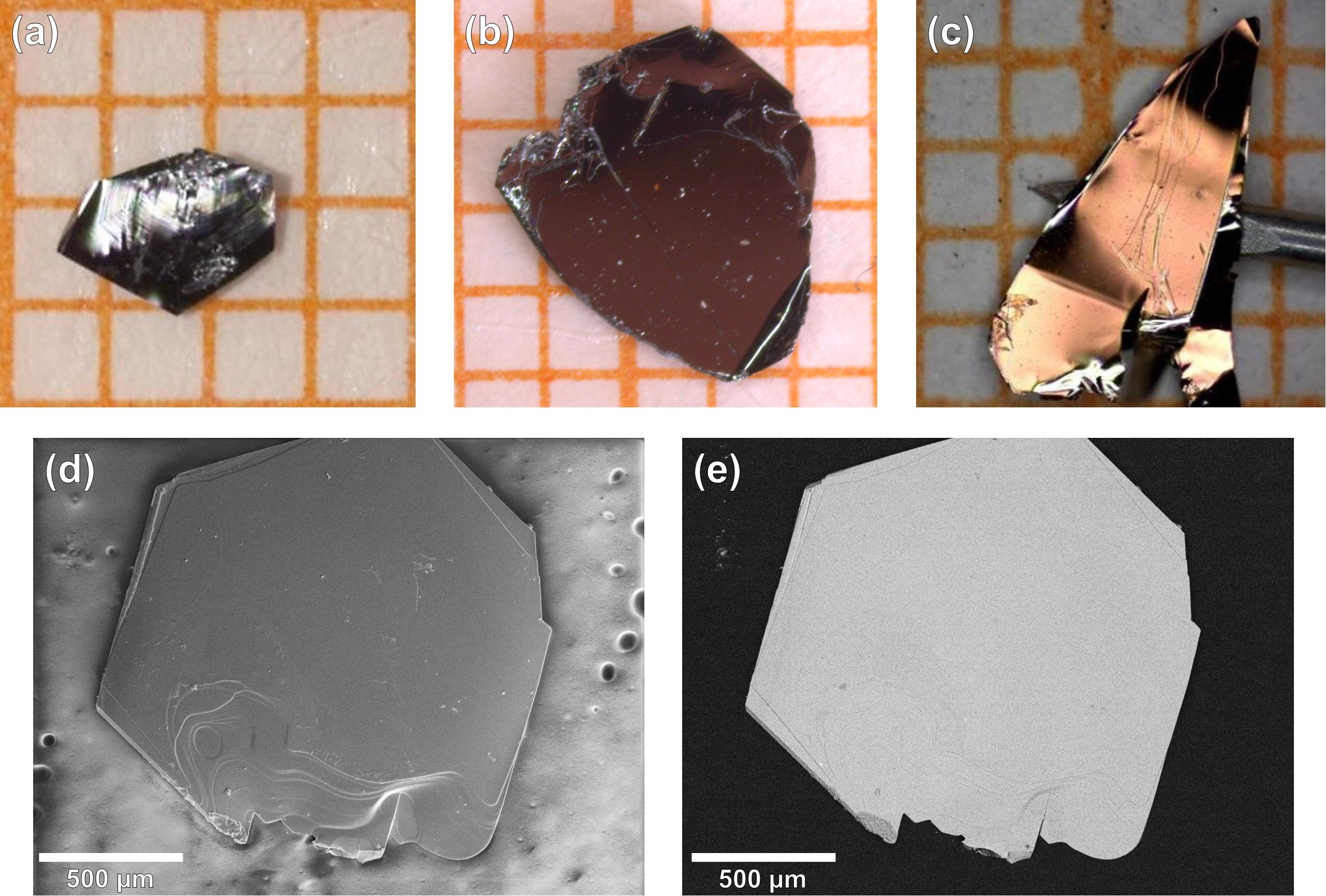}
\caption{As-grown single crystal of (a) Fe$_2$P$_2$S$_6$, (b) (Fe$_{0.7}$Ni$_{0.3}$)$_2$P$_2$S$_6$ and (c) Ni$_2$P$_2$S$_6$. An orange square in the background corresponds to 1\,mm\,$\times$\,1\,mm for scale. Electron microscopy images of a (Fe$_{0.7}$Ni$_{0.3}$)$_2$P$_2$S$_6$ crystal in topographic mode (SE detector) in (d) and in chemical contrast mode (BSE detector) in (e).}
\label{fig:crystal_images}
\end{figure}

Initially, the furnace was heated homogeneously to 710\,\gc\ with 100\,\gc/h. The charge side was kept at this temperature for 394.5\,h while the other side of the ampule, referred to as the sink side, was initially heated up to 760\,\gc\ with 100\,\gc/h, dwelled at this temperature for 24\,h and then cooled back to 710\,\gc\ with 1\,\gc/h. Then, an inverse transport gradient is formed, i.e., transport from the sink to the charge side, to clean the sink side of particles which stuck to the walls of the quartz ampule during filling. This ensures improved nucleation conditions in the following step. The sink side was cooled to 660\,\gc\ with 0.5\,\gc/h to slowly form the thermal transport gradient resulting in a controlled nucleation. Then, the ampule was dwelled with a transport gradient of 710\,\gc\ (charge) to 660\,\gc\ (sink) for 200\,h. After this the charge side was cooled to the sink temperature in 1\,h before both sides were furnace cooled to room temperature. Lustrous black plate-like crystals of Fe$_2$P$_2$S$_6$ of up to 5\,mm\,$\times$\,5\,mm\,$\times$\,200\,$\mu$m were obtained. An as-grown single crystal of Fe$_2$P$_2$S$_6$ is shown in Fig.~\ref{fig:crystal_images}(a). The crystals exhibit a layered morphology and a ductile nature. Furthermore, the crystals are easily exfoliated which is a typical feature of these layered vdW-materials.

\subsubsection{Ni$_2$P$_2$S$_6$}

Single crystals of Ni$_2$P$_2$S$_6$ were grown following the same general procedure as described for Fe$_2$P$_2$S$_6$, however, using a different growth temperature profile. A two-zone furnace was heated homogeneously to 750\,\gc\ with 100\,\gc/h. The charge side was kept at this temperature for 394.5\,h while the sink side was heated up to 800\,\gc\ with 100\,\gc/h, dwelled at this temperature for 24\,h and then cooled back to 750\,\gc\ with 1\,\gc/h. Subsequently, the sink side was cooled to 690\,\gc\ with 0.5\,\gc/h followed by a dwelling with this transport gradient (charge: 760\,\gc\ to sink: 690\,\gc) for 200\,h. Then, the charge side was cooled to the sink temperature in 1\,h before both sides were furnace cooled to room temperature. This temperature profile is adapted from Taylor~\textit{et~al.}~\cite{Taylor1973}. Shiny brown flake-like crystals of Ni$_2$P$_2$S$_6$ of up to 4\,mm\,$\times$\,3\,mm\,$\times$\,200\,$\mu$m were obtained. An exemplary as-grown crystal of Ni$_2$P$_2$S$_6$ is shown in Fig.~\ref{fig:crystal_images}~(c).

\subsubsection{(Fe$_{1-x}$Ni$_x$)$_2$P$_2$S$_6$}

Using the pre-reacted polycrystalline material as starting material and following the same procedure for the CVT growth as described before with the same temperature profile as for Ni$_2$P$_2$S$_6$, we obtained crystals of (Fe$_{1-x}$Ni$_x$)$_2$P$_2$S$_6$. However, the surface of some crystals obtained by this procedure was contaminated with byproducts which condensed during or after the CVT (most likely iodine, phosphorus and FeI$_2$). These surface impurities could be removed by washing the crystal shortly with water or acetone without any notable decomposition or structural changes of the main phase of the crystal. With these optimized conditions single crystals of (Fe$_{1-x}$Ni$_x$)$_2$P$_2$S$_6$ with a mass of up to 20\,mg were grown. A representative as grown single crystal of (Fe$_{0.7}$Ni$_{0.3}$)$_2$P$_2$S$_6$ is shown in Fig. \ref{fig:crystal_images} (b) together with its electron microscopic images in Figs. \ref{fig:crystal_images}(d) and (e).

\subsection{Methods}

\begin{figure*}[htb]
\includegraphics[width=\textwidth]{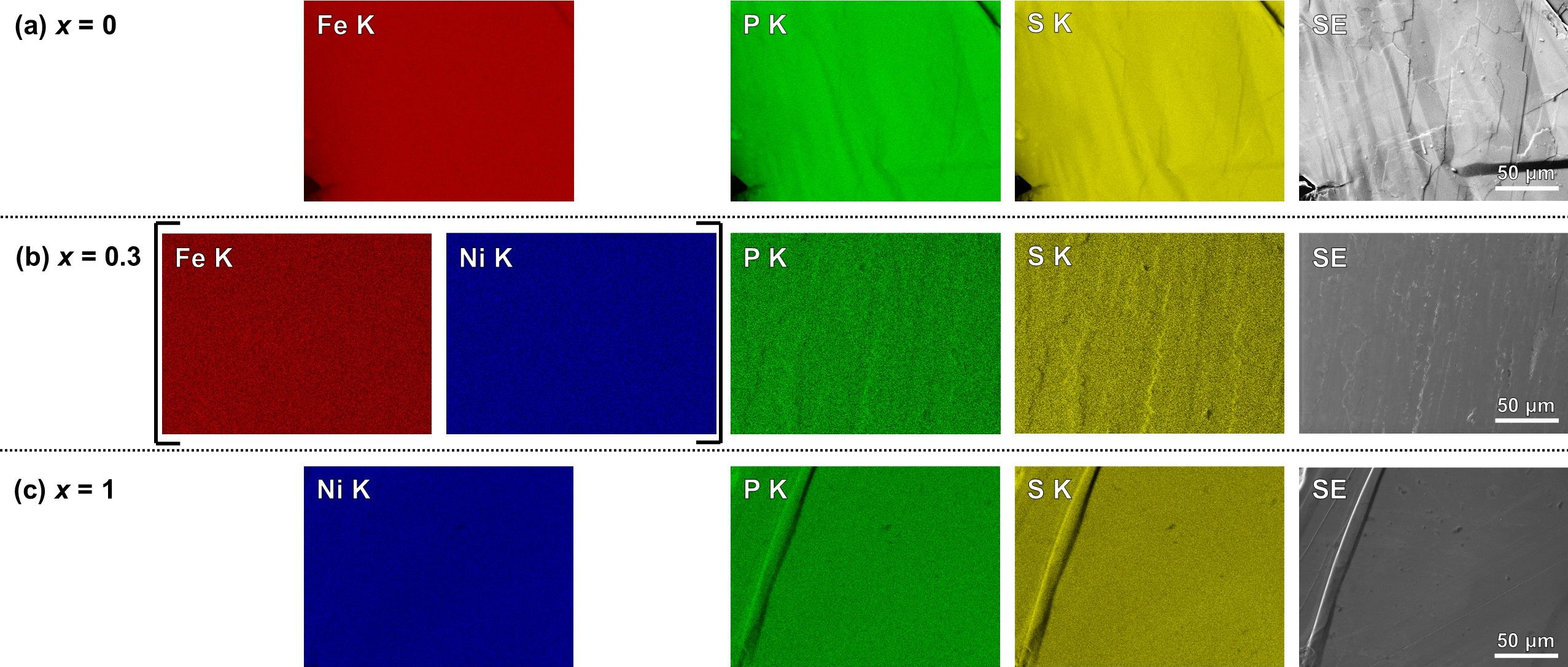}
\caption{Elemental maps and SEM(SE) image of a crystal of (a) Fe$_2$P$_2$S$_6$ ($x = 0$), (b) (Fe$_{0.7}$Ni$_{0.3}$)$_2$P$_2$S$_6$ ($x = 0.3$) and (c) Ni$_2$P$_2$S$_6$ ($x = 1$). The elemental maps are color-coded based on the relative intensity of the characteristic X-ray K line. Brighter regions corresponds to higher relative intensity.}
\label{fig:elemental_mapping}
\end{figure*}

All crystals were thoroughly characterized by several techniques. Single crystal X-ray diffraction (scXRD) was performed at room temperature on a Bruker-AXS KAPPA APEX II CCD diffractometer with graphite-monochromated Mo-K$_\alpha$ radiation (50\,kV, 30\,mA). The data collection consists of large $\Omega$ and $\phi$ scans of the reciprocal space. The frames were integrated with the Bruker SAINT software package~\footnote{SAINT(V8.30A), Bruker AXS Inc., Madison, Wisconsin,
USA (2017)} using a narrow-frame algorithm in APEX3~\footnote{Bruker, APEX3 v2018.1-0, Bruker AXS Inc., Madison,
Wisconsin, USA (2017)}. The data were corrected for absorption effects using a semiempirical method based on redundancy with the SADABS program~\cite{Krause2015}, developed for scaling and absorption corrections of area detector data. The space group determination, structural determination and refinement were performed using charge flipping with the Superflip algorithm \cite{Palatinus2007} within Jana2006~\cite{Petricek2014} and SHELXL~\cite{Sheldrick2007}. The crystal structures were refined with anisotropic displacement parameters for all atoms.

Powder X-ray diffraction (pXRD) was performed on pulverized crystals at room temperature on a STOE STADI laboratory diffractometer in transmission geometry with Cu-K$_{\alpha1}$ radiation from a curved Ge(111) single crystal monochromator and detected by a MYTHEN 1K 12.5$^\circ$-linear position sensitive detector manufactured by DECTRIS. Jana2006~\cite{Petricek2014} was used to analyze the pXRD diffraction pattern with the Le~Bail method.

A ZEISS EVO MA 10 scanning electron microscope (SEM) was used to capture electron microscopic images using either a secondary electron (SE) detector for topographic contrast or a backscattered electron (BSE) detector for chemical contrast. The same SEM device allowed for energy dispersive X-ray spectroscopy (EDX) measurements at an accelerating voltage of 30\,kV using an energy dispersive X-ray analyzer. EDX point and area scans were performed on crystals which were attached to the sample holder with carbon tape. To minimize sample drift and avoid charge-up effects during the long time measurements for EDX elemental maps, these crystals were embedded in an epoxy resin mixed with fine carbon powder to ensure conductivity. Samples were embedded such that they formed a clean surface with the surrounding resin. On this surface, a thin film (8\,nm) of gold was evaporated to further improve the sample conductivity.

The DC magnetization was measured as a function of temperature in an applied magnetic field of $\mu_0H = 1$\,T using a superconducting quantum interference device vibrating sample magnetometer (SQUID-VSM) from Quantum Design. All measurements were performed during heating after the sample was cooled in an applied field of 1\,T to 1.8\,K (field cooled; fc).

\section{Results}
\subsection{Scanning electron microscopy and compositional analysis}

From the SEM images using the topographic contrast mode (SE), crystals of all  Ni-substitution exhibit the typical features of layered systems, such as steps and terraces, \textit{e.g.}, as shown in Fig.~\ref{fig:crystal_images}(d) for (Fe$_{0.7}$Ni$_{0.3}$)$_2$P$_2$S$_6$. Using the chemical contrast mode (BSE), no changes in contrast are observed on the surface of the crystals indicating a homogeneous elemental composition. This is exemplarily shown in Fig.~\ref{fig:crystal_images}(e) for (Fe$_{0.7}$Ni$_{0.3}$)$_2$P$_2$S$_6$.

EDX was measured on at least 10 different spots on several crystals of each nominal degree of Ni-substitution, EDX mapping was performed on an area of approximately 150 $\times$ 200 $\mu$m$^2$ of one crystal for each Ni-substituted crystal. While the former measurements yield information about the homogeneity of the elemental composition of a crystal growth batch, the latter yields detailed insight in the local homogeneity of the elemental composition, complementary to the SEM(BSE) images. In fact, changes in the elemental composition which do not strongly affect the local mean atomic number, such as a slight change in the Fe:Ni ratio in (Fe$_{1-x}$Ni$_{x}$)$_2$P$_2$S$_6$, are expected to result in the same contrast in the SEM(BSE) image, while a notable shift in spectral weight in the corresponding EDX spectra may still be observed.

However for our (Fe$_{1-x}$Ni$_{x}$)$_2$P$_2$S$_6$ crystals, also the elemental maps obtained from EDX mapping indicate a homogeneous distribution of Fe, Ni, P and S without notable changes of the Fe:Ni ratio for all nominal degrees of Ni-substitution, supporting the conclusions from the SEM (BSE) images. The elemental maps for both parent compounds as well as for (Fe$_{0.7}$Ni$_{0.3}$)$_2$P$_2$S$_6$ are shown in Fig.~\ref{fig:elemental_mapping}. It should be noted, that the observed contrast changes in the elemental maps are purely attributed to surface features and not actually to compositional changes, as clearly seen by comparing the elemental maps to the SEM(SE) image of the same area. Furthermore, an EDX investigation on the cross section of a (Fe$_{0.7}$Ni$_{0.3}$)$_2$P$_2$S$_6$ crystal also revealed no compositional gradients along the stacking direction of the layers. Thus, the crystals of the (Fe$_{1-x}$Ni$_{x}$)$_2$P$_2$S$_6$ substitution series are homogeneous in composition.

\begin{table}[htb]
\begin{tabular}{lccc}
\hline \hline
\multirow{2}{*}{$x_\textrm{nom}$} & \multicolumn{2}{c}{Composition} & \multirow{2}{*}{$x_\textrm{exp}$} \\
 & Expected & Measured &  \\
\hline
0   & Fe$_{2}$P$_{2}$S$_{6}$          & Fe$_{2.07(3)}$P$_{2.03(1)}$S$_{5.90(3)}$              & 0    \\
0.3 & Fe$_{1.4}$Ni$_{0.6}$P$_{2}$S$_{6}$  & Fe$_{1.45(4)}$Ni$_{0.59(1)}$P$_{2.04(2)}$S$_{5.93(3)}$ & 0.29 \\
0.5 & Fe$_{1.0}$Ni$_{1.0}$P$_{2}$S$_{6}$ & Fe$_{1.11(3)}$Ni$_{0.96(3)}$P$_{2.02(1)}$S$_{5.91(6)}$ & 0.46 \\
0.7 & Fe$_{0.6}$Ni$_{1.4}$P$_{2}$S$_{6}$  & Fe$_{0.65(1)}$Ni$_{1.40(2)}$P$_{2.03(1)}$S$_{5.92(2)}$ & 0.68 \\
0.9 & Fe$_{0.2}$Ni$_{1.8}$P$_{2}$S$_{6}$  & Fe$_{0.23(3)}$Ni$_{1.87(8)}$P$_{2.02(2)}$S$_{5.87(8)}$ & 0.89 \\
1   & Ni$_{2}$P$_{2}$S$_{6}$          & Ni$_{2.05(1)}$P$_{2.03(1)}$S$_{5.92(1)}$              & 1    \\
\hline \hline
\end{tabular}
\caption{Expected and mean composition measured by EDX as well as the resulting estimated experimental Ni substitution for crystals with different nominal substitution. Standard deviations are given in parentheses.}
\label{tab:elemental_composition}
\end{table}

The mean elemental composition for the crystals of the (Fe$_{1-x}$Ni$_{x}$)$_2$P$_2$S$_6$ substitution series were obtained from multiple EDX spot measurements and are summarized in Tab.~\ref{tab:elemental_composition}. The elemental compositions are close to the expected composition of the $M_2$P$_2$S$_6$ phase and are homogeneous over multiple crystals of the same crystal growth batch as indicated by small standard deviations. Moreover, for all crystals the degree of Ni-substitution $x_\textrm{exp}$ is found in the range of the nominal value $x_\textrm{nom}$ considering a systematic uncertainty of this ratio of approximately 5\%.

\subsection{Structural analysis}

All scXRD patterns could be indexed in the same monoclinic space group $C2/m$ (No.~12). The atomic model proposed by Klingen~\textit{et al.} for Fe$_2$P$_2$S$_6$~\cite{Klingen1973} was sufficient to describe all obtained scXRD pattern with reasonable reliability. In this model, the transition element $4g$ Wyckoff sites are octahedrally surrounded by Sulfur sites (Wyckoff $4i$ \& $8j$). These octahedra are connected via the edges to form a honeycomb network in the $ab$ plane. In the void of each honeycomb, a P$_2$ dumbbell is located (Wyckoff $4i$) with the P-P bond being perpendicular to the $ab$ plane. These layers are interacting with each other solely by van der Waals forces and are stacked onto each other with an offset in the $a$ direction due to the monoclinic angle $\beta \neq 90^\circ$ between $a$ and $c$.

\begin{figure}[tb]
\includegraphics[width=\columnwidth]{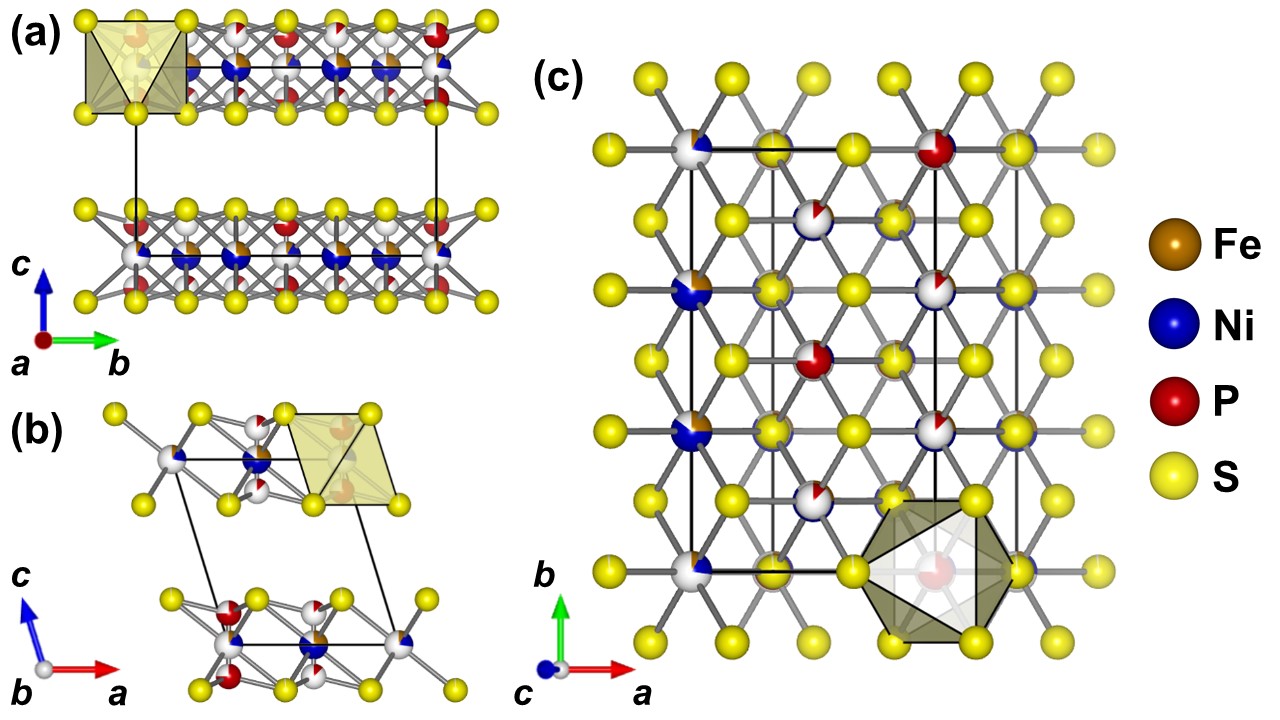}
\caption{Crystal structure of (Fe$_{1-x}$Ni$_x$)$_2$P$_2$S$_6$. Exemplary relative occupancies of the $T$ position by Fe and Ni as well as the disorder ratio are according to the structural parameters obtained from scXRD for (Fe$_{0.3}$Ni$_{0.7}$)$_2$P$_2$S$_6$ (see Tab.~\ref{tab:lattice_parameter_scXRD_pXRD}). (a) shows the $bc$ plane, (b) shows the $ac$ plane and (c) shows the $ab$ plane. Each S$_6$ octahedron either hosts a transition metal atom or a P$_2$ dumbbell in its center, as indicated by the yellow octahedron. Less than half of the occupied atomic sites correspond to the minority transition element 2a and P 8j sites. The graphical representation was prepared using VESTA\,3~\cite{KMomma2011}.}
\label{fig:crystal_structure}
\end{figure}

\begin{figure}[tb]
\includegraphics[width=\columnwidth]{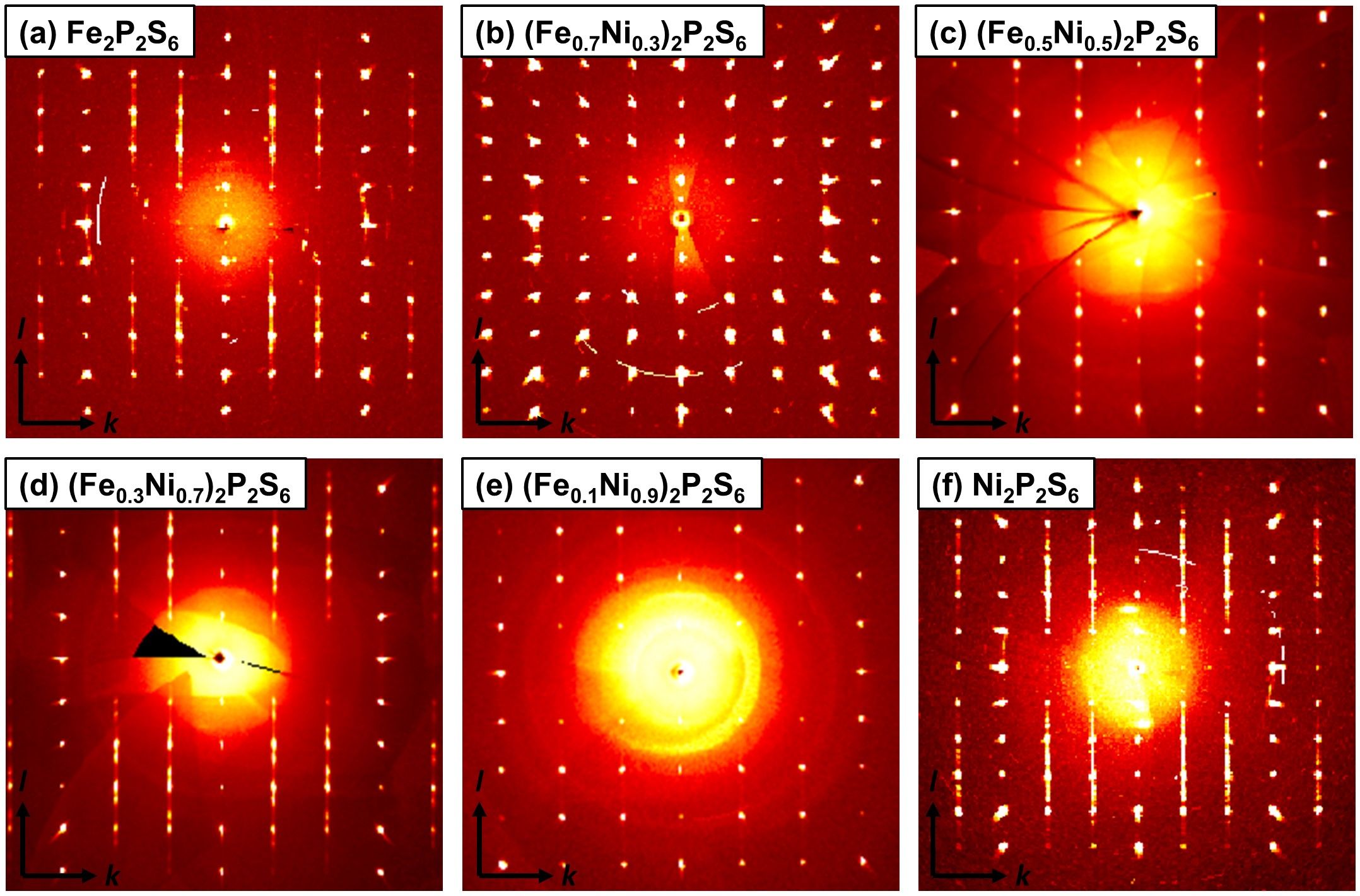}
\caption{
$0kl$ section of reciprocal space measured by scXRD on (Fe$_{1-x}$Ni$_{x}$)$_2$P$_2$S$_6$ with
(a) $x = 0$, (b) $x = 0.3$, (c) $x = 0.5$, (d) $x = 0.7$, (e) $x = 0.9$ and (f) $x = 1$. Significant streaking
in l-direction (perpendicular to the ab-planes) is observable in (a), (d) and (f).
}
\label{fig:scXRD_0kl}
\end{figure}

In extension of the model of Klingen~\textit{et al.}, introducing a site disorder of the transition element between majority $4g$ (Fe1 and Ni1) and minority $2a$ sites (Fe2 and Ni2) and of P between majority $4i$ (P1) and minority $8j$ sites (P2) as proposed by Ouvard~\textit{et al.} for Ni$_2$P$_2$S$_6$~\cite{Ouvrard1985}, improved the agreement between model and experiment even further. A disorder ratio $r_\textrm{disorder}$ is defined in Eq.~\ref{eq:disorder_ratio}, which describes the ratio between the majority and minority sites of transition element atoms and takes into account the different site multiplicities. The respective lattice parameters and the site disorder ratio $r_\textrm{disorder}$ based on the evaluation of the scXRD experiments are shortly summarized in Tab.~\ref{tab:lattice_parameter_scXRD_pXRD}~(top). A graphical illustration of the crystal structure of (Fe$_{1-x}$Ni$_x$)$_2$P$_2$S$_6$ including the site disorder is shown in Fig.~\ref{fig:crystal_structure}.

\begin{equation}
r_\textrm{disorder} = \dfrac{f_\textrm{Fe2} + f_\textrm{Ni2}}{2(f_\textrm{Fe1} + f_\textrm{Ni1}) + f_\textrm{Fe2} + f_\textrm{Ni2}}
\label{eq:disorder_ratio}
\end{equation}

\begin{table*}[tb]
\centering
\caption{
Summary of lattice parameters and volume obtained from modelling the scXRD and pXRD pattern. A selection of reliability values are shown as well.
}
\begin{tabular}{lcccccc}
\hline \hline
\multicolumn{7}{c}{scXRD} \\
\hline
 $x_{\textrm{nom}}$ & 0.0 & 0.3 & 0.5 & 0.7 & 0.9 & 1 \\
\hline
Crystal System & \multicolumn{6}{c}{Monoclinic} \\
Space Group & \multicolumn{6}{c}{$C2/m$ (No. 12)} \\
\textit{a} (\r{A}) & 5.9345(10) & 5.8544(2) & 5.8602(2) & 5.8671(2) & 5.8481(3) & 5.8165(7) \\
\textit{b} (\r{A}) & 10.2942(17) & 10.1487(2) & 10.1497(4) & 10.1658(3) & 10.1219(3) & 10.0737(12) \\
\textit{c} (\r{A}) & 6.7092(11) & 6.6563(2) & 6.6518(2) & 6.6663(2) & 6.6534(3) & 6.6213(8) \\
$\beta$ ($^\circ$) & 107.069(11) & 107.139(4) & 107.069(3) & 107.086(3) & 107.127(6) & 107.110(6) \\
Volume (\r{A}$^3$) & 391.82(11) & 377.92(2) & 378.22(2) & 380.06(2) & 376.38(3) & 370.80(10) \\
$r_\textrm{disorder}$ (\%) & 6 & 0 (fixed) & 2 & 14 & 2 & 4 \\
Goodness-Of-Fit & 1.13 & 2.30 & 1.61 & 2.55 & 1.93 & 1.28 \\
$R_\textrm{obs}$ (\%) & 1.70 & 2.32 & 3.50 & 2.67 & 2.93 & 2.17 \\
\hline
\multicolumn{7}{c}{pXRD (Le Bail)} \\
\hline
Crystal System & \multicolumn{6}{c}{Monoclinic} \\
Space Group & \multicolumn{6}{c}{$C2/m$ (No. 12)} \\
\textit{a} (\r{A}) & 5.9349(1) & 5.8919(7) & 5.8501(1) & 5.8513(2) & 5.8071(8) & 5.8224(2) \\
\textit{b} (\r{A}) & 10.2812(2) & 10.1809(12) & 10.1308(2) & 10.1337(3) & 10.0569(15) & 10.0831(3) \\
\textit{c} (\r{A}) & 6.7180(2) & 6.6779(9) & 6.6567(2) & 6.6571(3) & 6.6105(10) & 6.6332(2) \\
$\beta$ ($^\circ$) & 107.309(2) & 107.244(3) & 107.199(2) & 107.204(2) & 107.263(3) & 107.096(2) \\
Volume (\r{A}$^3$) & 391.35(2) & 382.57(8) & 376.87(2) & 377.07(2) & 368.67(9) & 372.22(2) \\
Goodness-Of-Fit & 1.69 & 2.04 & 2.31 & 1.97 & 1.97 & 2.77 \\
$R_\textrm{p}$ & 1.78 & 1.84 & 2.31 & 2.73 & 2.97 & 3.30 \\
$wR_\textrm{p}$ & 2.63 & 2.74 & 3.65 & 4.41 & 4.43 & 5.39 \\
\hline \hline
\end{tabular}
\label{tab:lattice_parameter_scXRD_pXRD}
\end{table*}

As already discussed by us in Ref.~\cite{Dioguardi2020} and by Goossens~\textit{et al.}~\cite{Goossens2011}, and by Lan\c{c}on~~\textit{et al.}~\cite{Lancon2018}, it is most likely that a high concentration of stacking faults leads to a significant displacement of electron density in layered $M_2$P$_2$S$_6$ systems. This displaced electron density could easily be falsely interpreted as the aforementioned site disorder in the structural refinement. As shown in Fig.~\ref{fig:scXRD_0kl}, a notable broadening of reflections is observed in the $l$ direction of the $0kl$ section of reciprocal space for (Fe$_{1-x}$Ni$_x$)$_2$P$_2$S$_6$ with $x_\textrm{Ni} =$ 0, 0.7, 1, while no significant broadening is observed for $x_\textrm{Ni} =$ 0.3, 0.5, 0.9. This effect in the scXRD pattern is indicative of (stacking) disorder perpendicular to the $ab$ planes rather than a large site disorder $r_{\textrm{disorder}}$ introduced above. This is further supported by angular dependent $^{31}$P-NMR spectra of our Ni$_2$P$_2$S$_6$ single crystals in the magnetically ordered state~\cite{Dioguardi2020}.

Although indications for stacking faults are clearly observed in the scXRD patterns, a crystal structure model, which explicitly takes these stacking faults into account, is lacking by now. Therefore, the model with site disorder was used to be able to obtain comparable structural parameters throughout the whole substitution series as it yields the best achievable agreement with the experiment. This is in line with the approach of Goossens~\textit{et al.}~\cite{Goossens2011} and Lan\c{c}on~\textit{et al.}~\cite{Lancon2018} for Ni$_2$P$_2$S$_6$. However, it should be emphasized, that the site disorder most likely is not real but rather an "artificial tool" to account for stacking faults in the refinement.

\begin{figure}[tb]
\includegraphics[width=\columnwidth]{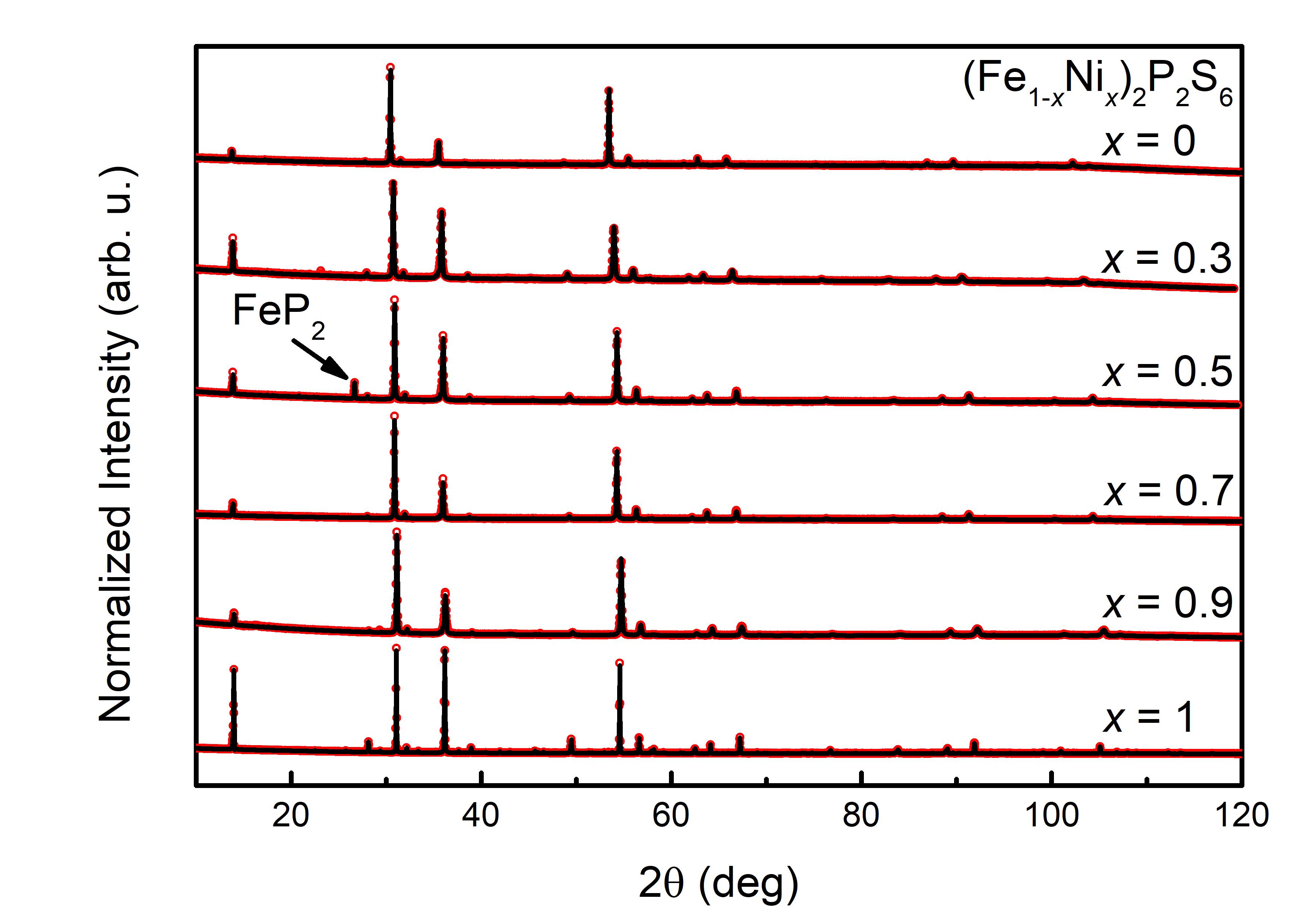}
\caption{
pXRD pattern of the members of the (Fe$_{1-x}$Ni$_x$)$_2$P$_2$S$_6$ substitution series from Cu-K$_{\alpha1}$ radiation ($\lambda = 1.5406$\,\r{A}) in red and corresponding Le~Bail fits in black. All pattern are normalized to the main reflection.
}
\label{fig:pXRD_summarized}
\end{figure}

As shown in Fig.~\ref{fig:pXRD_summarized}, all pXRD pattern are successfully indexed in the same monoclinic space group $C2/m$ (No.~12) as the scXRD pattern. Evidently, the aforementioned structural faults also affect the pXRD pattern, as an asymmetric shape of the $00l$ reflections is observed in our experiments. Additionally, significant altered relative intensities compared to an initial model are detected, which are attributed to the effect of preferred orientation. As pulverizing plate-like crystals yields plate-like rather than spherical powder particles, preferred orientation effects are commonly observed for layered compounds with weak interlayer interactions. However, such effects pose an additional challenge in the accurate modeling of the experimental pattern. Using the Le~Bail method, a reasonable description for the pattern profiles could be found and lattice parameters could be extracted. Corresponding lattice parameter are shown in Tab.~\ref{tab:lattice_parameter_scXRD_pXRD}~(bottom).

\begin{figure}[tb]
\includegraphics[width=\columnwidth]{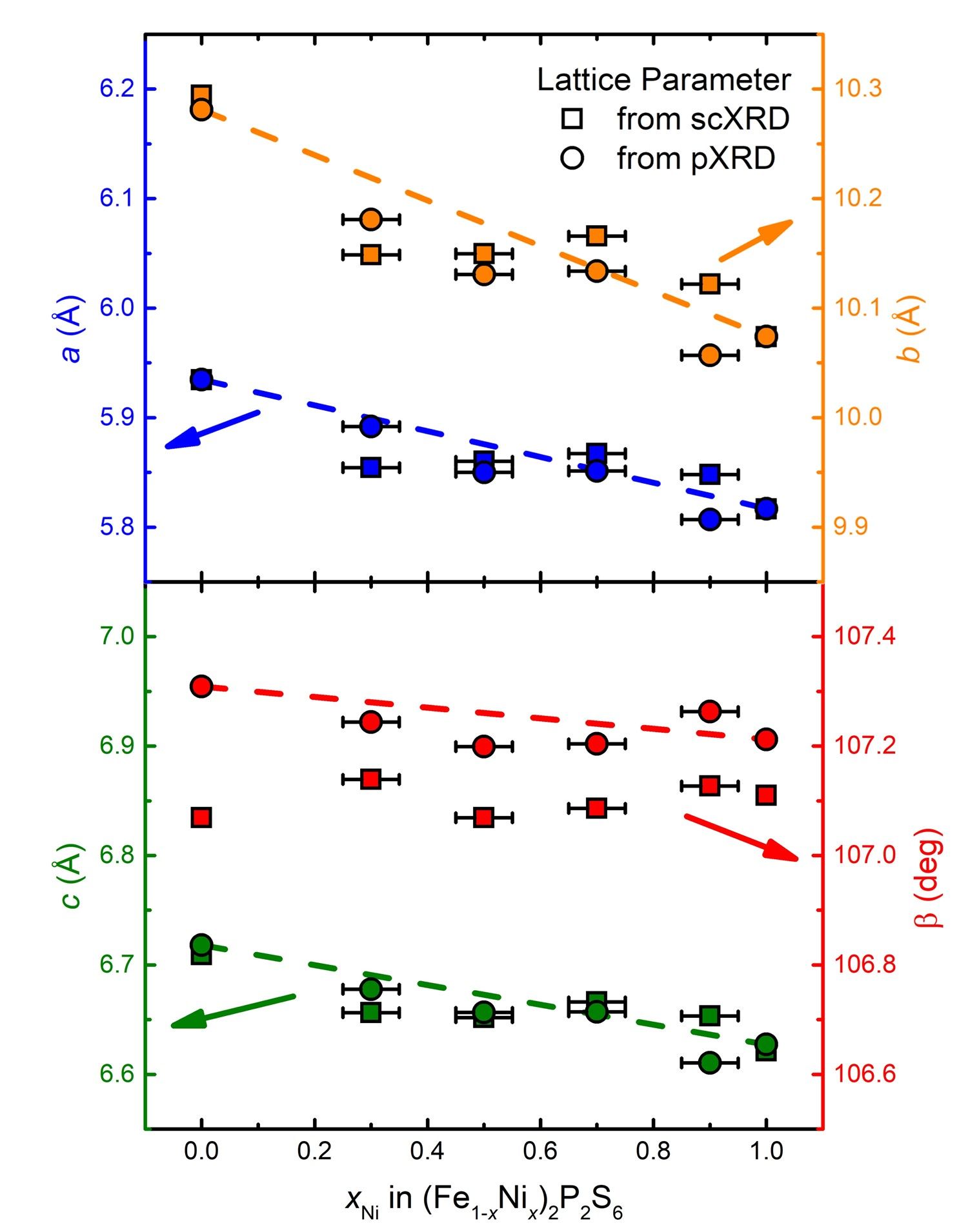}
\caption{
Evolution of the lattice parameters (top: $a$ and $b$; bottom: $c$ and $\beta$) of (Fe$_{1-x}$Ni$_x$)$_2$P$_2$S$_6$ as function of the degree of Ni substitution. Squares correspond to the values extracted from scXRD and circles show the values extracted from pXRD by Le Bail fits. The dashed lines illustrate an ideal linear evolution based on the lattice parameters of Fe$_2$P$_2$S$_6$ and Ni$_2$P$_2$S$_6$ from pXRD. Errorbars are smaller than the symbols.
}
\label{fig:evolution_lattice_parameter_cell_volume}
\end{figure}

The evolution of the lattice parameters $a$, $b$ and $c$ and the monoclinic angle $\beta$ extracted from pXRD and scXRD experiments is shown in Fig.~\ref{fig:evolution_lattice_parameter_cell_volume}. As expected from the ionic radii of Fe$^{2+}$ (0.78~\AA) in comparison to Ni$^{2+}$ (0.69~\AA)~\cite{RShannon1976}, the lattice parameters $a$, $b$ and $c$ shrink by increasing $x_\textrm{Ni}$ in (Fe$_{1-x}$Ni$_x$)$_2$P$_2$S$_6$. Overall a linear trend is observed, in agreement with Vegard's law~\cite{Vegard1921}. The monoclinic angle $\beta$ virtually stays constant as function of the degree of Ni substitution $x_\textrm{Ni}$. However, all these structural parameters exhibit small deviations from the general trend. It is likely that the structural faults in the real structure of (Fe$_{1-x}$Ni$_x$)$_2$P$_2$S$_6$ induce an additional inaccuracy in the determination of the lattice parameters, which results in the observed deviations. The investigations of the stacking faults by Transmission Electron Microscopy (TEM) is a separate on going investigation and beyond the scope of this present work. 

To conclude the structural analysis, both scXRD and pXRD confirm the formation of solid solution in (Fe$_{1-x}$Ni$_x$)$_2$P$_2$S$_6$. On the one hand, all scXRD pattern could be modeled using the same structural model only adjusting the Fe/Ni ratio. On the other hand, lattice parameters extracted from both scXRD and pXRD pattern overall follow Vegard's law.

\subsection{Magnetic analysis}

\begin{figure*}[tb]
\includegraphics[width=\textwidth]{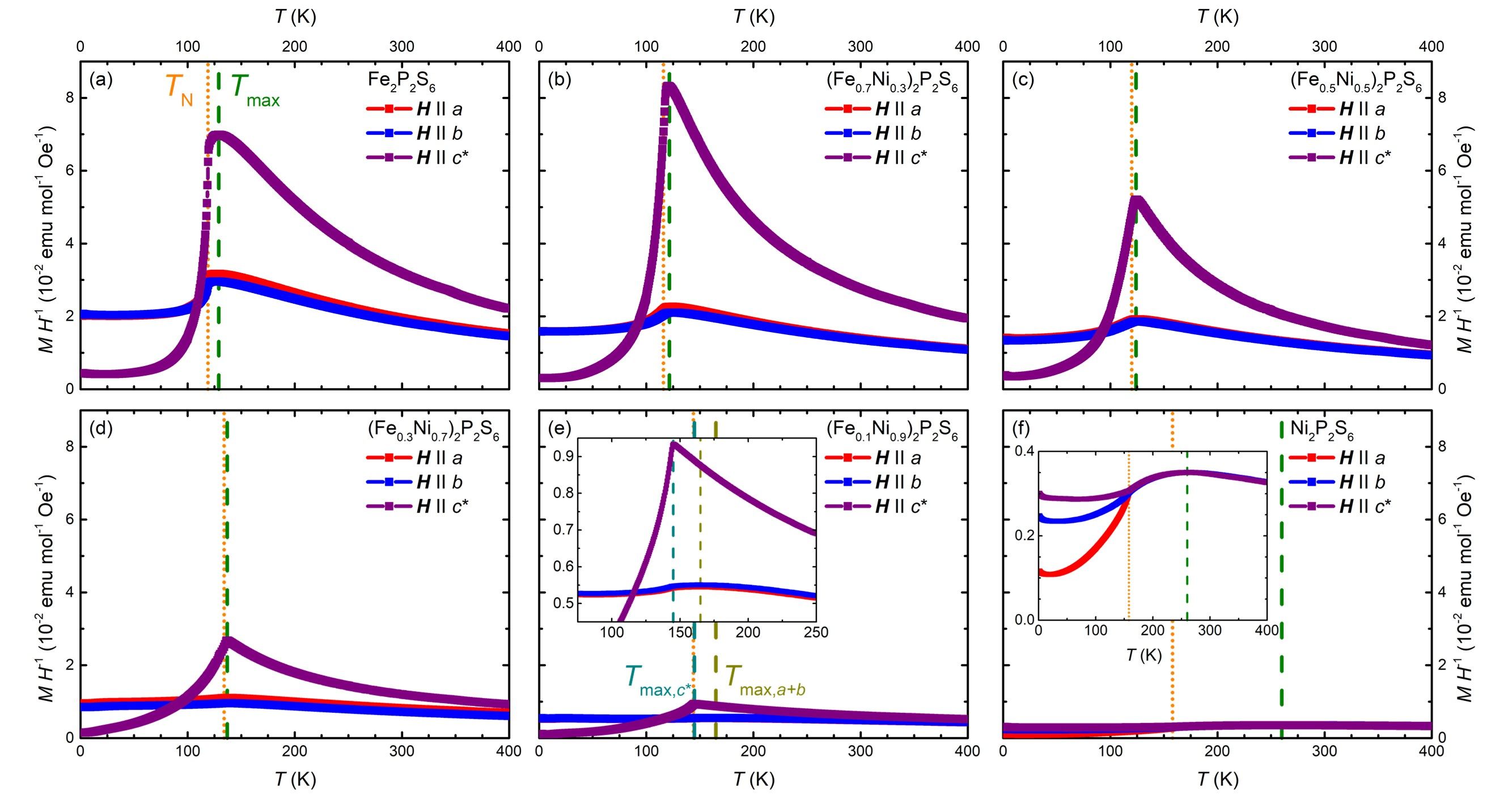}
\caption{Normalized magnetization as function of temperature $MH^{-1}(T)$ for a field of 1~T applied along three different crystallographic directions for (a) Fe$_2$P$_2$S$_6$, (b) (Fe$_{0.7}$Ni$_{0.3}$)$_2$P$_2$S$_6$, (c) (Fe$_{0.5}$Ni$_{0.5}$)$_2$P$_2$S$_6$, (d) (Fe$_{0.3}$Ni$_{0.7}$)$_2$P$_2$S$_6$, (e) (Fe$_{0.1}$Ni$_{0.9}$)$_2$P$_2$S$_6$ and (f) Ni$_2$P$_2$S$_6$. The $MH^{-1}$ axes have the same scale for best comparability. The insets in (e) and (f) show a zoomed-in view on the thermal evolution of $MH^{-1}$ of (Fe$_{0.1}$Ni$_{0.9}$)$_2$P$_2$S$_6$ and Ni$_2$P$_2$S$_6$, respectively. The orange dotted lines indicate the N\'{e}el temperature $T_\textrm{N}$ for each compound. The green dashed lines in (a), (b), (c), (d) and (f) denote the isotropic temperature of the maximum of $MH^{-1}(T)$ $T_\textrm{max}$. For (e) (Fe$_{0.1}$Ni$_{0.9}$)$_2$P$_2$S$_6$, $T_\textrm{max}$ is found to be anisotropic and accordingly the yellow line indicates $T_\textrm{max}$ for $\boldsymbol{H} \parallel a$ and $\boldsymbol{H} \parallel b$ while the turquoise line marks $T_\textrm{max}$ for $\boldsymbol{H} \parallel c\textrm{*}$.}
\label{fig:MH-1_T_evolution}
\end{figure*}

The normalized magnetization as function of temperature $MH^{-1}(T)$ is shown in Fig.~\ref{fig:MH-1_T_evolution} for (Fe$_{1-x}$Ni$_x$)$_2$P$_2$S$_6$ with (a) $x_\textrm{Ni} = 0$, (b) $x_\textrm{Ni} = 0.3$, (c) $x_\textrm{Ni} = 0.5$, (d) $x_\textrm{Ni} = 0.7$, (e) $x_\textrm{Ni} = 0.9$, (f) $x_\textrm{Ni} = 1$ with a field of $\mu_0H = 1$\,T applied parallel to three main crystallographic axis $a$, $b$, and $c^*$. The overall behavior of $MH^{-1}(T)$ is the same for all compositions and field directions. From 400~K towards lower temperatures, first an increase in $MH^{-1}$ is observed, followed by a decrease. Subsequently, a maximum of $MH^{-1}$ is obtained at a composition dependent temperature $T_\textrm{max}$. Below $T_\textrm{max}$, an inflection point is found at a composition dependent temperature, reflecting the onset of long-range magnetic order and thus defining the antiferromagnetic transition temperature $T_\textrm{N}$. Our observations are in agreement with existing literature for both parent compounds, where antiferromagnetic long-range order has been observed via different methods~\cite{PJoy1992,AWildes2015} including nuclear magnetic resonance and M{\"o}{\ss}bauer spectroscopy ~\cite{Dioguardi2020,CBerthier1978,Jernberg1984}, and neutron diffraction~\cite{AWildes2015,KKurosawa1983,KRule2007,DLancon2016}.

Although all members of the (Fe$_{1-x}$Ni$_x$)$_2$P$_2$S$_6$ substitution series share the same characteristic features in the temperature dependent magnetization, the overall shape notable differs, as \textit{e.g.} observed by comparing both parent compounds. For Fe$_2$P$_2$S$_6$, $T_\textrm{N}$ and $T_\textrm{max}$ are relatively close to each other with a sharp maximum in $M/H$, while for Ni$_2$P$_2$S$_6$, $T_\textrm{max}$ is much higher than $T_\textrm{N}$ with a broad maximum at high temperature. The former resembles a rather textbook-like shape of an antiferromagnetic--paramagnetic transition~\cite{Blundell2001}. The latter exhibits a shape which is observed for many other low-dimensional antiferromagnets (\textit{e.g.}, MnTiO$_3$~\cite{JAkimitsu1970} and K$_2$NiF$_4$~\cite{RBirgeneau1969}) in which the broad maximum is typically related to strong low-dimensional magnetic correlations above the long-range magnetic ordering temperature \cite{Lueken1999}. For the intermediate members of (Fe$_{1-x}$Ni$_x$)$_2$P$_2$S$_6$ with $x_\textrm{Ni} =$ 0.3, 0.5, 0.7, $T_\textrm{N}$ and $T_\textrm{max}$ are close together, such that the overall shape of $MH^{-1}(T)$ around the maximum resembles the one of Fe$_2$P$_2$S$_6$ rather than of Ni$_2$P$_2$S$_6$. Remarkably, (Fe$_{0.1}$Ni$_{0.9}$)$_2$P$_2$S$_6$ is unique in this substitution series, as it is the only compound for which $T_\textrm{max}$ undergoes an orientation dependence. As shown in the inset of Fig.~\ref{fig:MH-1_T_evolution}(e), the anisotropic behavior of $T_\textrm{max}$ in (Fe$_{0.1}$Ni$_{0.9}$)$_2$P$_2$S$_6$ manifests in a sharp maximum ($T_\textrm{max}$ being close to $T_\textrm{N}$) for $\boldsymbol{H} \parallel c^*$ and a notably broader maximum ($T_\textrm{max}$ being significantly higher than $T_\textrm{N}$) for $\boldsymbol{H} \parallel a$ and $\boldsymbol{H} \parallel b$. While these observations apparently follow the necessary changes for the evolution from an anisotropic Ising magnet (Fe$_2$P$_2$S$_6$) towards an in-plane Heisenberg magnet (Ni$_2$P$_2$S$_6$), it is interesting to see that the first physical quantity showing abrupt changes upon the series is $T_{\textrm{max}}$, which is related to the (low-dimensional) magnetic correlations and not 3D order in the system.

\begin{figure}[tb]
\includegraphics[width=\columnwidth]{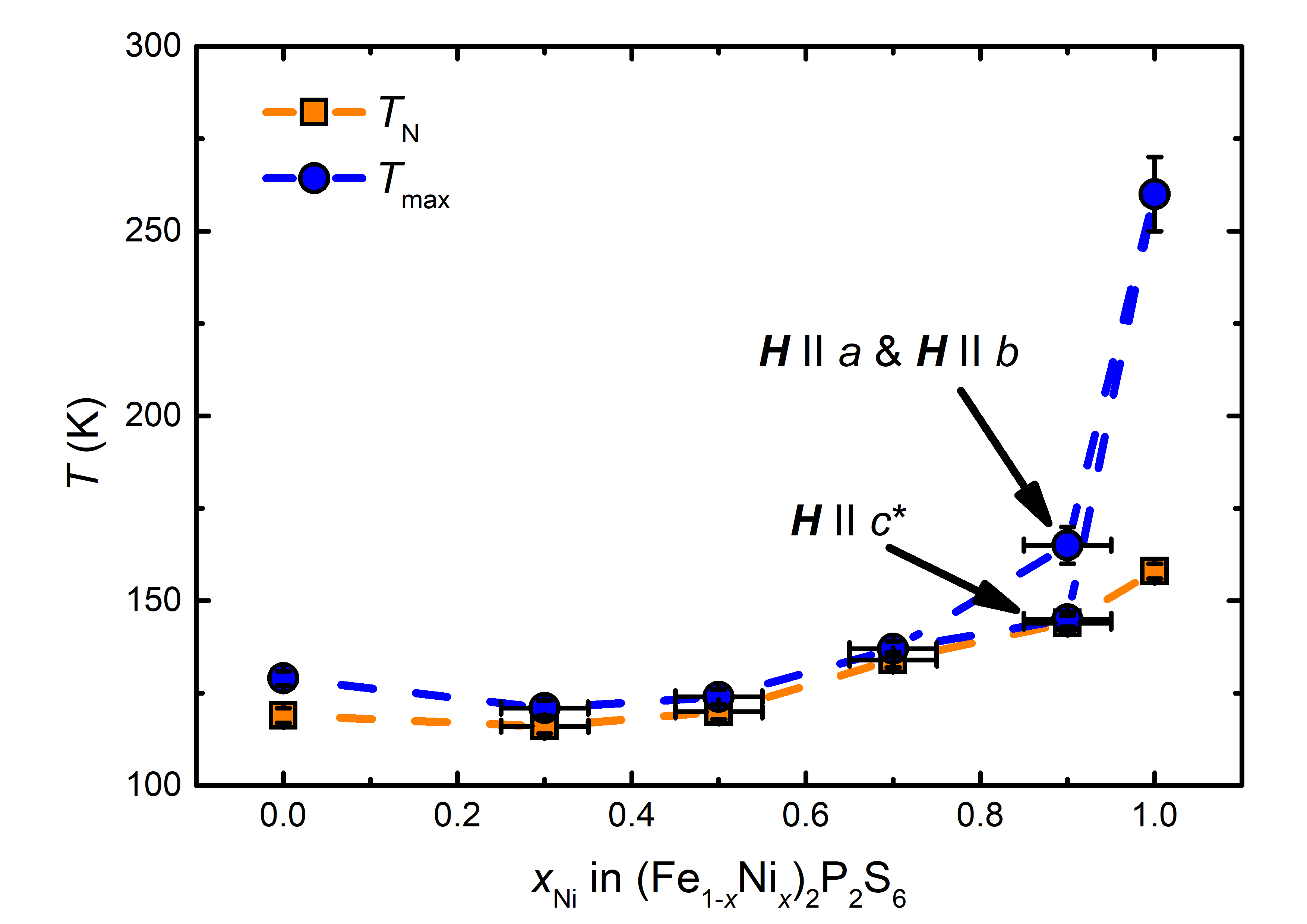}
\caption{Evolution of the N\'eel temperature $T_\textrm{N}$ and the temperature of the maximum in $MH^{-1}(T)$ $T_\textrm{max}$ as function of the degree of Ni substitution in (Fe$_{1-x}$Ni$_x$)$_2$P$_2$S$_6$ single crystals.}
\label{fig:TN_Tmax}
\end{figure}

The evolution of $T_\textrm{N}$ and $T_\textrm{max}$ as function of the degree of Ni substitution is shown in Fig.~\ref{fig:TN_Tmax}. Our observations are in qualitative agreement with the results of Rao and Rachaudhuri based on polycrystalline samples of the (Fe$_{1-x}$Ni$_x$)$_2$P$_2$S$_6$ substitution series~\cite{RRao1992}, where, however, due to the polycrystalline nature of their samples, the influence of the direction of the magnetic field at high degrees of Ni substitution could not be observed. $T_\textrm{N}$ exhibits a gradual increase over the full substitution series from Fe$_2$P$_2$S$_6$ to Ni$_2$P$_2$S$_6$. When the magnetic field is applied $\parallel c^*$, $T_\textrm{max}$ follows the gradual evolution of $T_\textrm{N}$ up to $x_\textrm{Ni} = 0.9$ and then increases rapidly between $x_\textrm{Ni} = 0.9$ and $x_\textrm{Ni} = 1$. Yet, when the field is applied $\parallel a$ or $\parallel b$, the evolution of $T_\textrm{max}$ appears to be more gradual, with $T_{\textrm{max}}$ deviating from the trend of $T_\textrm{N}$ already above $x_{\textrm{Ni}}$ = 0.7, evidencing a highly anisotropic behavior for samples with $x_\textrm{Ni}$ $>$ 0.7. This is in line with the (expected) strengthening of magnetic correlations in the ab plane towards the in-plane Heisenberg magnet Ni$_2$P$_2$S$_6$.

The single-crystalline nature of our samples allows to comment further on the orientation dependence of the normalized magnetization with respect to the alignment in a magnetic field,(\textit{i.e.}, the magnetic anisotropy) and its evolution as function of Ni substitution in (Fe$_{1-x}$Ni$_x$)$_2$P$_2$S$_6$. In the following, we will split our observations for temperatures below and above the magnetic long-range ordered state. Below $T_\textrm{N}$, for Fe$_2$P$_2$S$_6$ we find $MH^{-1}_{\parallel c^*} < MH^{-1}_{\parallel a} = MH^{-1}_{\parallel b}$, which is the expected behavior for an Ising-like antiferromagnet with $c^*$ being the magnetic easy axis in agreement with literature~\cite{PJoy1992,DLancon2016}. All intermediate compounds of (Fe$_{1-x}$Ni$_x$)$_2$P$_2$S$_6$ ($x_\textrm{Ni} =$ 0.3, 0.5, 0.7, 0.9) exhibit the same anisotropic magnetization in the magnetically ordered state. Only for Ni$_2$P$_2$S$_6$, a distinctly different behavior with $MH^{-1}_{\parallel a} < MH^{-1}_{\parallel b} < MH^{-1}_{\parallel c^*}$ is observed, which \textit{per se} is in agreement with literature~\cite{PJoy1992,AWildes2015}. Note that similar to $T_\textrm{max}$, which is related to (dynamic) magnetic correlations discussed above, also the (static) long-range ordered state features an abrupt change of the magnetic anisotropic behavior for $x_\textrm{Ni}$ $>$ 0.9.

Above the magnetic ordering temperature $T~>~T_\textrm{N}$, a peculiar orientation dependence of the normalized magnetization is observed for Fe$_2$P$_2$S$_6$ with $MH^{-1}_{\parallel c^*}~>~MH^{-1}_{\parallel a}~=~MH^{-1}_{\parallel b}$ in agreement with literature~\cite{Jernberg1984,PJoy1992}. Joy and Vasudevan~\cite{PJoy1992} attributed this magnetic anisotropic behavior in the paramagnetic state of Fe$_2$P$_2$S$_6$ to an anisotropy of the Land\'{e} factor $g$ and thus to details of the crystallographic electric field splitting scheme. A comparable orientation dependence of the magnetization above the magnetic ordering temperature is also observed for all intermediate compounds of (Fe$_{1-x}$Ni$_x$)$_2$P$_2$S$_6$ ($x_\textrm{Ni} =$ 0.3, 0.5, 0.7, 0.9). Only for Ni$_2$P$_2$S$_6$, an isotropic magnetic behavior is detected in the paramagnetic state, which is independent of the magnetic field direction in line with a weakly anisotropic Heisenberg system.

\section{Conclusions and Summary}

We report the successful single crystal growth of the layered quasi-2D system (Fe$_{1-x}$Ni$_x$)$_2$P$_2$S$_6$ with $0 \leq x \leq 1$ using the chemical vapor transport growth technique from polycrystalline stoichiometric precursors with iodine as transport agent. Utilization of pre-reacted polycrystalline material as starting material has proven crucial for the crystal growth to obtain crystals with the desired degree of Ni substitution and a sufficient elemental homogeneity. EDX mappings illustrate the homogeneity of the composition with in the limit of the technique. Single crystal XRD patterns for crystals of all degrees of Ni substitution are well described by the same atomic model proposed by Ouvard \textit{et al.}~\cite{Ouvrard1985}. Lattice parameters extracted from scXRD and pXRD patterns exhibit an overall linear evolution in agreement with Vegard's law~\cite{Vegard1921}.

Our temperature dependent magnetization measurements on the single-crystalline members of the (Fe$_{1-x}$Ni$_x$)$_2$P$_2$S$_6$ substitution series exhibit an antiferromagnetic ground state throughout the series. The N\'{e}el temperature $T_\textrm{N}$, evolves gradually over the substitution series.  For (Fe$_{0.1}$Ni$_{0.9}$)$_2$P$_2$S$_6$, a peculiar anisotropy is detected for $T_\textrm{max}$ manifesting in a sharp maximum for $\boldsymbol{H} \parallel c^*$ and a broad maximum for $\boldsymbol{H} \parallel a$ and $\boldsymbol{H} \parallel b$. Similarly, the magnetic anisotropic behavior for the next member $x_\textrm{Ni}$ = 1, i.e., Ni$_2$P$_2$S$_6$, clearly differs from the rest of the (Fe$_{1-x}$Ni$_x$)$_2$P$_2$S$_6$ substitution series. In the paramagnetic state above $T_\textrm{N}$, Ni$_2$P$_2$S$_6$ is the only investigated member of the (Fe$_{1-x}$Ni$_x$)$_2$P$_2$S$_6$ substitution series with an isotropic magnetization. For all other members, $MH^{-1}_{\parallel c^*} > MH^{-1}_{\parallel a} = MH^{-1}_{\parallel b}$ is observed. The magnetic anisotropy as well as the temperature of the correlation maximum $T_\textrm{max}$ do not evolve gradually throughout the substitution series but rather show two distinctly different regimes, one for (Fe$_{1-x}$Ni$_x$)$_2$P$_2$S$_6$ up to $x_\textrm{Ni} =$ 0.9 and another one for Ni$_2$P$_2$S$_6$. The former regime is in agreement with an Ising-like anisotropy, while the magnetic anisotropy of Ni$_2$P$_2$S$_6$ is best described by a weakly anisotropic Heisenberg model or the XXZ model according to literature~\cite{PJoy1992,DLancon2016,Kim2019}.

The origin of the observed abrupt evolution of the magnetic anisotropy around $x_\textrm{Ni}$ may be subject to future investigations, for example if it can be continuously tuned from Ising-like to XXZ-like in the narrow regime between (Fe$_{0.1}$Ni$_{0.9}$)$_2$P$_2$S$_6$ and Ni$_2$P$_2$S$_6$. Also, it will be interesting to compare the evolution of the magnetic behavior and anisotropy of (Fe$_{1-x}$Ni$_x$)$_2$P$_2$S$_6$ in detail to closely related systems, such as (Mn$_{1-x}$Fe$_x$)$_2$P$_2$S$_6$~\cite{Masubuchi2008} and (Mn$_{1-x}$Ni$_x$)$_2$P$_2$S$_6$~\cite{Shemerliuk2021}. Finally, we emphasize that crystals grown in this work may also allow to study the substitution dependence of electronic properties within this series. Recently, Ni$_2$P$_2$S$_6$ and other compounds of the $M_2$P$_2$S$_6$ family were investigated regarding their performance in field effect transistors (FETs) \cite{Jenjeti2018,Kumar2019}. Our single crystals of the (Fe$_{1-x}$Ni$_x$)$_2$P$_2$S$_6$ substitution series may allow to manufacture several FETs and study the performance of these devices as function of the degree of substitution in the future.

\section{Acknowledgements}
This work is supported by the Deutsche Forschungsgemeinschaft (DFG) via Grant No.~DFG~A.S~523\textbackslash4-1. M.S. acknowledges financial support \textit{via} Grant No. STU 695\textbackslash1-1. A.U.B.W. and B.B. acknowledge financial support from the DFG through SFB~1143~(project-id 247310070) and the W\"urzburg-Dresden Cluster of Excellence on Complexity and Topology in Quantum Matter -- \textit{ct.qmat} (EXC 2147, project-id 390858490).

\bibliography{NiFePS3_Literature}

\begin{thebibliography}{48}%
\makeatletter
\providecommand \@ifxundefined [1]{%
 \@ifx{#1\undefined}
}%
\providecommand \@ifnum [1]{%
 \ifnum #1\expandafter \@firstoftwo
 \else \expandafter \@secondoftwo
 \fi
}%
\providecommand \@ifx [1]{%
 \ifx #1\expandafter \@firstoftwo
 \else \expandafter \@secondoftwo
 \fi
}%
\providecommand \natexlab [1]{#1}%
\providecommand \enquote  [1]{``#1''}%
\providecommand \bibnamefont  [1]{#1}%
\providecommand \bibfnamefont [1]{#1}%
\providecommand \citenamefont [1]{#1}%
\providecommand \href@noop [0]{\@secondoftwo}%
\providecommand \href [0]{\begingroup \@sanitize@url \@href}%
\providecommand \@href[1]{\@@startlink{#1}\@@href}%
\providecommand \@@href[1]{\endgroup#1\@@endlink}%
\providecommand \@sanitize@url [0]{\catcode `\\12\catcode `\$12\catcode
  `\&12\catcode `\#12\catcode `\^12\catcode `\_12\catcode `\%12\relax}%
\providecommand \@@startlink[1]{}%
\providecommand \@@endlink[0]{}%
\providecommand \url  [0]{\begingroup\@sanitize@url \@url }%
\providecommand \@url [1]{\endgroup\@href {#1}{\urlprefix }}%
\providecommand \urlprefix  [0]{URL }%
\providecommand \Eprint [0]{\href }%
\providecommand \doibase [0]{https://doi.org/}%
\providecommand \selectlanguage [0]{\@gobble}%
\providecommand \bibinfo  [0]{\@secondoftwo}%
\providecommand \bibfield  [0]{\@secondoftwo}%
\providecommand \translation [1]{[#1]}%
\providecommand \BibitemOpen [0]{}%
\providecommand \bibitemStop [0]{}%
\providecommand \bibitemNoStop [0]{.\EOS\space}%
\providecommand \EOS [0]{\spacefactor3000\relax}%
\providecommand \BibitemShut  [1]{\csname bibitem#1\endcsname}%
\let\auto@bib@innerbib\@empty
%</preamble>
\bibitem [{\citenamefont {Huang}\ \emph {et~al.}(2017)\citenamefont {Huang},
  \citenamefont {Clark}, \citenamefont {Navarro-Moratalla}, \citenamefont
  {Klein}, \citenamefont {Cheng}, \citenamefont {Seyler}, \citenamefont
  {Zhong}, \citenamefont {Schmidgall}, \citenamefont {McGuire}, \citenamefont
  {Cobden}, \citenamefont {Yao}, \citenamefont {Xiao}, \citenamefont
  {Jarillo-Herrero},\ and\ \citenamefont {Xu}}]{Huang2017}%
  \BibitemOpen
  \bibfield  {author} {\bibinfo {author} {\bibfnamefont {B.}~\bibnamefont
  {Huang}}, \bibinfo {author} {\bibfnamefont {G.}~\bibnamefont {Clark}},
  \bibinfo {author} {\bibfnamefont {E.}~\bibnamefont {Navarro-Moratalla}},
  \bibinfo {author} {\bibfnamefont {D.~R.}\ \bibnamefont {Klein}}, \bibinfo
  {author} {\bibfnamefont {R.}~\bibnamefont {Cheng}}, \bibinfo {author}
  {\bibfnamefont {K.~L.}\ \bibnamefont {Seyler}}, \bibinfo {author}
  {\bibfnamefont {D.}~\bibnamefont {Zhong}}, \bibinfo {author} {\bibfnamefont
  {E.}~\bibnamefont {Schmidgall}}, \bibinfo {author} {\bibfnamefont {M.~A.}\
  \bibnamefont {McGuire}}, \bibinfo {author} {\bibfnamefont {D.~H.}\
  \bibnamefont {Cobden}}, \bibinfo {author} {\bibfnamefont {W.}~\bibnamefont
  {Yao}}, \bibinfo {author} {\bibfnamefont {D.}~\bibnamefont {Xiao}}, \bibinfo
  {author} {\bibfnamefont {P.}~\bibnamefont {Jarillo-Herrero}},\ and\ \bibinfo
  {author} {\bibfnamefont {X.}~\bibnamefont {Xu}},\ }\bibfield  {title}
  {\bibinfo {title} {Layer-dependent ferromagnetism in a van der waals crystal
  down to the monolayer limit},\ }\href {https://doi.org/10.1038/nature22391}
  {\bibfield  {journal} {\bibinfo  {journal} {Nature}\ }\textbf {\bibinfo
  {volume} {546}},\ \bibinfo {pages} {270} (\bibinfo {year}
  {2017})}\BibitemShut {NoStop}%
\bibitem [{\citenamefont {Gong}\ \emph {et~al.}(2017)\citenamefont {Gong},
  \citenamefont {Li}, \citenamefont {Li}, \citenamefont {Ji}, \citenamefont
  {Stern}, \citenamefont {Xia}, \citenamefont {Cao}, \citenamefont {Bao},
  \citenamefont {Wang}, \citenamefont {Wang}, \citenamefont {Qiu},
  \citenamefont {Cava}, \citenamefont {Louie}, \citenamefont {Xia},\ and\
  \citenamefont {Zhang}}]{CGong2017}%
  \BibitemOpen
  \bibfield  {author} {\bibinfo {author} {\bibfnamefont {C.}~\bibnamefont
  {Gong}}, \bibinfo {author} {\bibfnamefont {L.}~\bibnamefont {Li}}, \bibinfo
  {author} {\bibfnamefont {Z.}~\bibnamefont {Li}}, \bibinfo {author}
  {\bibfnamefont {H.}~\bibnamefont {Ji}}, \bibinfo {author} {\bibfnamefont
  {A.}~\bibnamefont {Stern}}, \bibinfo {author} {\bibfnamefont
  {Y.}~\bibnamefont {Xia}}, \bibinfo {author} {\bibfnamefont {T.}~\bibnamefont
  {Cao}}, \bibinfo {author} {\bibfnamefont {W.}~\bibnamefont {Bao}}, \bibinfo
  {author} {\bibfnamefont {C.}~\bibnamefont {Wang}}, \bibinfo {author}
  {\bibfnamefont {Y.}~\bibnamefont {Wang}}, \bibinfo {author} {\bibfnamefont
  {Z.~Q.}\ \bibnamefont {Qiu}}, \bibinfo {author} {\bibfnamefont {R.~J.}\
  \bibnamefont {Cava}}, \bibinfo {author} {\bibfnamefont {S.~G.}\ \bibnamefont
  {Louie}}, \bibinfo {author} {\bibfnamefont {J.}~\bibnamefont {Xia}},\ and\
  \bibinfo {author} {\bibfnamefont {X.}~\bibnamefont {Zhang}},\ }\bibfield
  {title} {\bibinfo {title} {Discovery of intrinsic ferromagnetism in
  two-dimensional van der waals crystals},\ }\href
  {https://doi.org/10.1038/nature22060} {\bibfield  {journal} {\bibinfo
  {journal} {Nature}\ }\textbf {\bibinfo {volume} {546}},\ \bibinfo {pages}
  {265} (\bibinfo {year} {2017})}\BibitemShut {NoStop}%
\bibitem [{Nat(2018)}]{NatNano2018}%
  \BibitemOpen
  \bibfield  {title} {\bibinfo {title} {{2D} magnetism gets hot},\ }\href
  {https://doi.org/10.1038/s41565-018-0128-9} {\bibfield  {journal} {\bibinfo
  {journal} {Nature Nanotechnology}\ }\textbf {\bibinfo {volume} {13}},\
  \bibinfo {pages} {269} (\bibinfo {year} {2018})}\BibitemShut {NoStop}%
\bibitem [{\citenamefont {Wang}\ \emph
  {et~al.}(2018{\natexlab{a}})\citenamefont {Wang}, \citenamefont {Sapkota},
  \citenamefont {Taniguchi}, \citenamefont {Watanabe}, \citenamefont
  {Mandrus},\ and\ \citenamefont {Morpurgo}}]{Wang2018}%
  \BibitemOpen
  \bibfield  {author} {\bibinfo {author} {\bibfnamefont {Z.}~\bibnamefont
  {Wang}}, \bibinfo {author} {\bibfnamefont {D.}~\bibnamefont {Sapkota}},
  \bibinfo {author} {\bibfnamefont {T.}~\bibnamefont {Taniguchi}}, \bibinfo
  {author} {\bibfnamefont {K.}~\bibnamefont {Watanabe}}, \bibinfo {author}
  {\bibfnamefont {D.}~\bibnamefont {Mandrus}},\ and\ \bibinfo {author}
  {\bibfnamefont {A.~F.}\ \bibnamefont {Morpurgo}},\ }\bibfield  {title}
  {\bibinfo {title} {Tunneling spin valves based on
  {Fe}$_{3}${GeTe}$_{2}$/{hBN}/{Fe}$_{3}${GeTe}$_{2}$ van der waals
  heterostructures},\ }\href {https://doi.org/10.1021/acs.nanolett.8b01278}
  {\bibfield  {journal} {\bibinfo  {journal} {Nano Letters}\ }\textbf {\bibinfo
  {volume} {18}},\ \bibinfo {pages} {4303} (\bibinfo {year}
  {2018}{\natexlab{a}})}\BibitemShut {NoStop}%
\bibitem [{\citenamefont {Gibertini}\ \emph {et~al.}(2019)\citenamefont
  {Gibertini}, \citenamefont {Koperski}, \citenamefont {Morpurgo},\ and\
  \citenamefont {Novoselov}}]{Gibertini2019}%
  \BibitemOpen
  \bibfield  {author} {\bibinfo {author} {\bibfnamefont {M.}~\bibnamefont
  {Gibertini}}, \bibinfo {author} {\bibfnamefont {M.}~\bibnamefont {Koperski}},
  \bibinfo {author} {\bibfnamefont {A.~F.}\ \bibnamefont {Morpurgo}},\ and\
  \bibinfo {author} {\bibfnamefont {K.~S.}\ \bibnamefont {Novoselov}},\
  }\bibfield  {title} {\bibinfo {title} {Magnetic {2D} materials and
  heterostructures},\ }\href {https://doi.org/10.1038/s41565-019-0438-6}
  {\bibfield  {journal} {\bibinfo  {journal} {Nature Nanotechnology}\ }\textbf
  {\bibinfo {volume} {14}},\ \bibinfo {pages} {408} (\bibinfo {year}
  {2019})}\BibitemShut {NoStop}%
\bibitem [{\citenamefont {Zhong}\ \emph {et~al.}(2017)\citenamefont {Zhong},
  \citenamefont {Seyler}, \citenamefont {Linpeng}, \citenamefont {Cheng},
  \citenamefont {Sivadas}, \citenamefont {Huang}, \citenamefont {Schmidgall},
  \citenamefont {Taniguchi}, \citenamefont {Watanabe}, \citenamefont {McGuire},
  \citenamefont {Yao}, \citenamefont {Xiao}, \citenamefont {Fu},\ and\
  \citenamefont {Xu}}]{Zhong2017}%
  \BibitemOpen
  \bibfield  {author} {\bibinfo {author} {\bibfnamefont {D.}~\bibnamefont
  {Zhong}}, \bibinfo {author} {\bibfnamefont {K.~L.}\ \bibnamefont {Seyler}},
  \bibinfo {author} {\bibfnamefont {X.}~\bibnamefont {Linpeng}}, \bibinfo
  {author} {\bibfnamefont {R.}~\bibnamefont {Cheng}}, \bibinfo {author}
  {\bibfnamefont {N.}~\bibnamefont {Sivadas}}, \bibinfo {author} {\bibfnamefont
  {B.}~\bibnamefont {Huang}}, \bibinfo {author} {\bibfnamefont
  {E.}~\bibnamefont {Schmidgall}}, \bibinfo {author} {\bibfnamefont
  {T.}~\bibnamefont {Taniguchi}}, \bibinfo {author} {\bibfnamefont
  {K.}~\bibnamefont {Watanabe}}, \bibinfo {author} {\bibfnamefont {M.~A.}\
  \bibnamefont {McGuire}}, \bibinfo {author} {\bibfnamefont {W.}~\bibnamefont
  {Yao}}, \bibinfo {author} {\bibfnamefont {D.}~\bibnamefont {Xiao}}, \bibinfo
  {author} {\bibfnamefont {K.-M.~C.}\ \bibnamefont {Fu}},\ and\ \bibinfo
  {author} {\bibfnamefont {X.}~\bibnamefont {Xu}},\ }\bibfield  {title}
  {\bibinfo {title} {Van der waals engineering of ferromagnetic semiconductor
  heterostructures for spin and valleytronics},\ }\href
  {https://doi.org/10.1126/sciadv.1603113} {\bibfield  {journal} {\bibinfo
  {journal} {Science Advances}\ }\textbf {\bibinfo {volume} {3}},\ \bibinfo
  {pages} {e1603113} (\bibinfo {year} {2017})}\BibitemShut {NoStop}%
\bibitem [{\citenamefont {Song}\ \emph {et~al.}(2019)\citenamefont {Song},
  \citenamefont {Tu}, \citenamefont {Carnahan}, \citenamefont {Cai},
  \citenamefont {Taniguchi}, \citenamefont {Watanabe}, \citenamefont {McGuire},
  \citenamefont {Cobden}, \citenamefont {Xiao}, \citenamefont {Yao},\ and\
  \citenamefont {Xu}}]{Song2019}%
  \BibitemOpen
  \bibfield  {author} {\bibinfo {author} {\bibfnamefont {T.}~\bibnamefont
  {Song}}, \bibinfo {author} {\bibfnamefont {M.~W.-Y.}\ \bibnamefont {Tu}},
  \bibinfo {author} {\bibfnamefont {C.}~\bibnamefont {Carnahan}}, \bibinfo
  {author} {\bibfnamefont {X.}~\bibnamefont {Cai}}, \bibinfo {author}
  {\bibfnamefont {T.}~\bibnamefont {Taniguchi}}, \bibinfo {author}
  {\bibfnamefont {K.}~\bibnamefont {Watanabe}}, \bibinfo {author}
  {\bibfnamefont {M.~A.}\ \bibnamefont {McGuire}}, \bibinfo {author}
  {\bibfnamefont {D.~H.}\ \bibnamefont {Cobden}}, \bibinfo {author}
  {\bibfnamefont {D.}~\bibnamefont {Xiao}}, \bibinfo {author} {\bibfnamefont
  {W.}~\bibnamefont {Yao}},\ and\ \bibinfo {author} {\bibfnamefont
  {X.}~\bibnamefont {Xu}},\ }\bibfield  {title} {\bibinfo {title} {Voltage
  control of a van der waals spin-filter magnetic tunnel junction},\ }\href
  {https://doi.org/10.1021/acs.nanolett.8b04160} {\bibfield  {journal}
  {\bibinfo  {journal} {Nano Letters}\ }\textbf {\bibinfo {volume} {19}},\
  \bibinfo {pages} {915} (\bibinfo {year} {2019})}\BibitemShut {NoStop}%
\bibitem [{\citenamefont {Wang}\ \emph
  {et~al.}(2018{\natexlab{b}})\citenamefont {Wang}, \citenamefont
  {Guti{\'{e}}rrez-Lezama}, \citenamefont {Ubrig}, \citenamefont {Kroner},
  \citenamefont {Gibertini}, \citenamefont {Taniguchi}, \citenamefont
  {Watanabe}, \citenamefont {Imamo{\u{g}}lu}, \citenamefont {Giannini},\ and\
  \citenamefont {Morpurgo}}]{Wang2018a}%
  \BibitemOpen
  \bibfield  {author} {\bibinfo {author} {\bibfnamefont {Z.}~\bibnamefont
  {Wang}}, \bibinfo {author} {\bibfnamefont {I.}~\bibnamefont
  {Guti{\'{e}}rrez-Lezama}}, \bibinfo {author} {\bibfnamefont {N.}~\bibnamefont
  {Ubrig}}, \bibinfo {author} {\bibfnamefont {M.}~\bibnamefont {Kroner}},
  \bibinfo {author} {\bibfnamefont {M.}~\bibnamefont {Gibertini}}, \bibinfo
  {author} {\bibfnamefont {T.}~\bibnamefont {Taniguchi}}, \bibinfo {author}
  {\bibfnamefont {K.}~\bibnamefont {Watanabe}}, \bibinfo {author}
  {\bibfnamefont {A.}~\bibnamefont {Imamo{\u{g}}lu}}, \bibinfo {author}
  {\bibfnamefont {E.}~\bibnamefont {Giannini}},\ and\ \bibinfo {author}
  {\bibfnamefont {A.~F.}\ \bibnamefont {Morpurgo}},\ }\bibfield  {title}
  {\bibinfo {title} {Very large tunneling magnetoresistance in layered magnetic
  semiconductor {CrI}$_{3}$},\ }\href
  {https://doi.org/10.1038/s41467-018-04953-8} {\bibfield  {journal} {\bibinfo
  {journal} {Nature Communications}\ }\textbf {\bibinfo {volume} {9}},\
  \bibinfo {pages} {2516} (\bibinfo {year} {2018}{\natexlab{b}})}\BibitemShut
  {NoStop}%
\bibitem [{\citenamefont {Samarth}(2017)}]{Samarth2017}%
  \BibitemOpen
  \bibfield  {author} {\bibinfo {author} {\bibfnamefont {N.}~\bibnamefont
  {Samarth}},\ }\bibfield  {title} {\bibinfo {title} {Magnetism in flatland},\
  }\href {https://doi.org/10.1038/546216a} {\bibfield  {journal} {\bibinfo
  {journal} {Nature}\ }\textbf {\bibinfo {volume} {546}},\ \bibinfo {pages}
  {216} (\bibinfo {year} {2017})}\BibitemShut {NoStop}%
\bibitem [{\citenamefont {Mermin}\ and\ \citenamefont
  {Wagner}(1966)}]{Mermin1966}%
  \BibitemOpen
  \bibfield  {author} {\bibinfo {author} {\bibfnamefont {N.~D.}\ \bibnamefont
  {Mermin}}\ and\ \bibinfo {author} {\bibfnamefont {H.}~\bibnamefont
  {Wagner}},\ }\bibfield  {title} {\bibinfo {title} {Absence of ferromagnetism
  or antiferromagnetism in one- or two-dimensional isotropic heisenberg
  models},\ }\href {https://doi.org/10.1103/physrevlett.17.1133} {\bibfield
  {journal} {\bibinfo  {journal} {Physical Review Letters}\ }\textbf {\bibinfo
  {volume} {17}},\ \bibinfo {pages} {1133} (\bibinfo {year}
  {1966})}\BibitemShut {NoStop}%
\bibitem [{\citenamefont {Kosterlitz}\ and\ \citenamefont
  {Thouless}(1973)}]{Kosterlitz1973}%
  \BibitemOpen
  \bibfield  {author} {\bibinfo {author} {\bibfnamefont {J.~M.}\ \bibnamefont
  {Kosterlitz}}\ and\ \bibinfo {author} {\bibfnamefont {D.~J.}\ \bibnamefont
  {Thouless}},\ }\bibfield  {title} {\bibinfo {title} {Ordering, metastability
  and phase transitions in two-dimensional systems},\ }\href
  {https://doi.org/10.1088/0022-3719/6/7/010} {\bibfield  {journal} {\bibinfo
  {journal} {Journal of Physics C: Solid State Physics}\ }\textbf {\bibinfo
  {volume} {6}},\ \bibinfo {pages} {1181} (\bibinfo {year} {1973})}\BibitemShut
  {NoStop}%
\bibitem [{\citenamefont {Berezinsky}(1970)}]{Berezinsky1970}%
  \BibitemOpen
  \bibfield  {author} {\bibinfo {author} {\bibfnamefont {V.~L.}\ \bibnamefont
  {Berezinsky}},\ }\bibfield  {title} {\bibinfo {title} {Destruction of long
  range order in one-dimensional and two-dimensional systems having a
  continuous symmetry group. {I.} {C}lassical systems},\ }\href@noop {}
  {\bibfield  {journal} {\bibinfo  {journal} {Zh. Eksp. Teor. Fiz.}\ }\textbf
  {\bibinfo {volume} {32}},\ \bibinfo {pages} {493} (\bibinfo {year}
  {1970})}\BibitemShut {NoStop}%
\bibitem [{\citenamefont {Brec}(1986)}]{Brec1986}%
  \BibitemOpen
  \bibfield  {author} {\bibinfo {author} {\bibfnamefont {R.}~\bibnamefont
  {Brec}},\ }\bibfield  {title} {\bibinfo {title} {Review on structural and
  chemical properties of transition metal phosphorous trisulfides
  {MPS}$_{3}$},\ }\href {https://doi.org/10.1016/0167-2738(86)90055-x}
  {\bibfield  {journal} {\bibinfo  {journal} {Solid State Ionics}\ }\textbf
  {\bibinfo {volume} {22}},\ \bibinfo {pages} {3} (\bibinfo {year}
  {1986})}\BibitemShut {NoStop}%
\bibitem [{\citenamefont {Joy}\ and\ \citenamefont
  {Vasudevan}(1992)}]{PJoy1992}%
  \BibitemOpen
  \bibfield  {author} {\bibinfo {author} {\bibfnamefont {P.~A.}\ \bibnamefont
  {Joy}}\ and\ \bibinfo {author} {\bibfnamefont {S.}~\bibnamefont
  {Vasudevan}},\ }\bibfield  {title} {\bibinfo {title} {Magnetism in the
  layered transition-metal thiophosphates {MPS}$_{3}$ ({M} = {Mn}, {Fe}, and
  {Ni})},\ }\href {https://doi.org/10.1103/physrevb.46.5425} {\bibfield
  {journal} {\bibinfo  {journal} {Physical Review B}\ }\textbf {\bibinfo
  {volume} {46}},\ \bibinfo {pages} {5425} (\bibinfo {year}
  {1992})}\BibitemShut {NoStop}%
\bibitem [{\citenamefont {Wildes}\ \emph {et~al.}(1998)\citenamefont {Wildes},
  \citenamefont {Roessli}, \citenamefont {Lebech},\ and\ \citenamefont
  {Godfrey}}]{Wildes1998}%
  \BibitemOpen
  \bibfield  {author} {\bibinfo {author} {\bibfnamefont {A.~R.}\ \bibnamefont
  {Wildes}}, \bibinfo {author} {\bibfnamefont {B.}~\bibnamefont {Roessli}},
  \bibinfo {author} {\bibfnamefont {B.}~\bibnamefont {Lebech}},\ and\ \bibinfo
  {author} {\bibfnamefont {K.~W.}\ \bibnamefont {Godfrey}},\ }\bibfield
  {title} {\bibinfo {title} {Spin waves and the critical behaviour of the
  magnetization in {MnPS}$_{3}$},\ }\href
  {https://doi.org/10.1088/0953-8984/10/28/020} {\bibfield  {journal} {\bibinfo
   {journal} {Journal of Physics: Condensed Matter}\ }\textbf {\bibinfo
  {volume} {10}},\ \bibinfo {pages} {6417} (\bibinfo {year}
  {1998})}\BibitemShut {NoStop}%
\bibitem [{\citenamefont {Lan{\c{c}}on}\ \emph {et~al.}(2016)\citenamefont
  {Lan{\c{c}}on}, \citenamefont {Walker}, \citenamefont {Ressouche},
  \citenamefont {Ouladdiaf}, \citenamefont {Rule}, \citenamefont {McIntyre},
  \citenamefont {Hicks}, \citenamefont {R{\o}nnow},\ and\ \citenamefont
  {Wildes}}]{DLancon2016}%
  \BibitemOpen
  \bibfield  {author} {\bibinfo {author} {\bibfnamefont {D.}~\bibnamefont
  {Lan{\c{c}}on}}, \bibinfo {author} {\bibfnamefont {H.~C.}\ \bibnamefont
  {Walker}}, \bibinfo {author} {\bibfnamefont {E.}~\bibnamefont {Ressouche}},
  \bibinfo {author} {\bibfnamefont {B.}~\bibnamefont {Ouladdiaf}}, \bibinfo
  {author} {\bibfnamefont {K.~C.}\ \bibnamefont {Rule}}, \bibinfo {author}
  {\bibfnamefont {G.~J.}\ \bibnamefont {McIntyre}}, \bibinfo {author}
  {\bibfnamefont {T.~J.}\ \bibnamefont {Hicks}}, \bibinfo {author}
  {\bibfnamefont {H.~M.}\ \bibnamefont {R{\o}nnow}},\ and\ \bibinfo {author}
  {\bibfnamefont {A.~R.}\ \bibnamefont {Wildes}},\ }\bibfield  {title}
  {\bibinfo {title} {Magnetic structure and magnon dynamics of the
  quasi-two-dimensional antiferromagnet {FePS}$_{3}$},\ }\href
  {https://doi.org/10.1103/physrevb.94.214407} {\bibfield  {journal} {\bibinfo
  {journal} {Physical Review B}\ }\textbf {\bibinfo {volume} {94}},\ \bibinfo
  {pages} {214407} (\bibinfo {year} {2016})}\BibitemShut {NoStop}%
\bibitem [{\citenamefont {Lan{\c{c}}on}\ \emph {et~al.}(2018)\citenamefont
  {Lan{\c{c}}on}, \citenamefont {Ewings}, \citenamefont {Guidi}, \citenamefont
  {Formisano},\ and\ \citenamefont {Wildes}}]{Lancon2018}%
  \BibitemOpen
  \bibfield  {author} {\bibinfo {author} {\bibfnamefont {D.}~\bibnamefont
  {Lan{\c{c}}on}}, \bibinfo {author} {\bibfnamefont {R.~A.}\ \bibnamefont
  {Ewings}}, \bibinfo {author} {\bibfnamefont {T.}~\bibnamefont {Guidi}},
  \bibinfo {author} {\bibfnamefont {F.}~\bibnamefont {Formisano}},\ and\
  \bibinfo {author} {\bibfnamefont {A.~R.}\ \bibnamefont {Wildes}},\ }\bibfield
   {title} {\bibinfo {title} {Magnetic exchange parameters and anisotropy of
  the quasi-two-dimensional antiferromagnet {NiPS}$_{3}$},\ }\href
  {https://doi.org/10.1103/physrevb.98.134414} {\bibfield  {journal} {\bibinfo
  {journal} {Physical Review B}\ }\textbf {\bibinfo {volume} {98}},\ \bibinfo
  {pages} {134414} (\bibinfo {year} {2018})}\BibitemShut {NoStop}%
\bibitem [{\citenamefont {Kim}\ \emph {et~al.}(2019)\citenamefont {Kim},
  \citenamefont {Lim}, \citenamefont {Lee}, \citenamefont {Lee}, \citenamefont
  {Kim}, \citenamefont {Park}, \citenamefont {Jeon}, \citenamefont {Park},
  \citenamefont {Park},\ and\ \citenamefont {Cheong}}]{Kim2019}%
  \BibitemOpen
  \bibfield  {author} {\bibinfo {author} {\bibfnamefont {K.}~\bibnamefont
  {Kim}}, \bibinfo {author} {\bibfnamefont {S.~Y.}\ \bibnamefont {Lim}},
  \bibinfo {author} {\bibfnamefont {J.-U.}\ \bibnamefont {Lee}}, \bibinfo
  {author} {\bibfnamefont {S.}~\bibnamefont {Lee}}, \bibinfo {author}
  {\bibfnamefont {T.~Y.}\ \bibnamefont {Kim}}, \bibinfo {author} {\bibfnamefont
  {K.}~\bibnamefont {Park}}, \bibinfo {author} {\bibfnamefont {G.~S.}\
  \bibnamefont {Jeon}}, \bibinfo {author} {\bibfnamefont {C.-H.}\ \bibnamefont
  {Park}}, \bibinfo {author} {\bibfnamefont {J.-G.}\ \bibnamefont {Park}},\
  and\ \bibinfo {author} {\bibfnamefont {H.}~\bibnamefont {Cheong}},\
  }\bibfield  {title} {\bibinfo {title} {Suppression of magnetic ordering in
  {XXZ}-type antiferromagnetic monolayer {NiPS}$_{3}$},\ }\href
  {https://doi.org/10.1038/s41467-018-08284-6} {\bibfield  {journal} {\bibinfo
  {journal} {Nature Communications}\ }\textbf {\bibinfo {volume} {10}},\
  \bibinfo {pages} {345} (\bibinfo {year} {2019})}\BibitemShut {NoStop}%
\bibitem [{\citenamefont {Lee}\ \emph {et~al.}(2016)\citenamefont {Lee},
  \citenamefont {Lee}, \citenamefont {Ryoo}, \citenamefont {Kang},
  \citenamefont {Kim}, \citenamefont {Kim}, \citenamefont {Park}, \citenamefont
  {Park},\ and\ \citenamefont {Cheong}}]{Lee2016}%
  \BibitemOpen
  \bibfield  {author} {\bibinfo {author} {\bibfnamefont {J.-U.}\ \bibnamefont
  {Lee}}, \bibinfo {author} {\bibfnamefont {S.}~\bibnamefont {Lee}}, \bibinfo
  {author} {\bibfnamefont {J.~H.}\ \bibnamefont {Ryoo}}, \bibinfo {author}
  {\bibfnamefont {S.}~\bibnamefont {Kang}}, \bibinfo {author} {\bibfnamefont
  {T.~Y.}\ \bibnamefont {Kim}}, \bibinfo {author} {\bibfnamefont
  {P.}~\bibnamefont {Kim}}, \bibinfo {author} {\bibfnamefont {C.-H.}\
  \bibnamefont {Park}}, \bibinfo {author} {\bibfnamefont {J.-G.}\ \bibnamefont
  {Park}},\ and\ \bibinfo {author} {\bibfnamefont {H.}~\bibnamefont {Cheong}},\
  }\bibfield  {title} {\bibinfo {title} {Ising-type magnetic ordering in
  atomically thin {FePS}$_{3}$},\ }\href
  {https://doi.org/10.1021/acs.nanolett.6b03052} {\bibfield  {journal}
  {\bibinfo  {journal} {Nano Letters}\ }\textbf {\bibinfo {volume} {16}},\
  \bibinfo {pages} {7433} (\bibinfo {year} {2016})}\BibitemShut {NoStop}%
\bibitem [{\citenamefont {Masubuchi}\ \emph {et~al.}(2008)\citenamefont
  {Masubuchi}, \citenamefont {Hoya}, \citenamefont {Watanabe}, \citenamefont
  {Takahashi}, \citenamefont {Ban}, \citenamefont {Ohkubo}, \citenamefont
  {Takase},\ and\ \citenamefont {Takano}}]{Masubuchi2008}%
  \BibitemOpen
  \bibfield  {author} {\bibinfo {author} {\bibfnamefont {T.}~\bibnamefont
  {Masubuchi}}, \bibinfo {author} {\bibfnamefont {H.}~\bibnamefont {Hoya}},
  \bibinfo {author} {\bibfnamefont {T.}~\bibnamefont {Watanabe}}, \bibinfo
  {author} {\bibfnamefont {Y.}~\bibnamefont {Takahashi}}, \bibinfo {author}
  {\bibfnamefont {S.}~\bibnamefont {Ban}}, \bibinfo {author} {\bibfnamefont
  {N.}~\bibnamefont {Ohkubo}}, \bibinfo {author} {\bibfnamefont
  {K.}~\bibnamefont {Takase}},\ and\ \bibinfo {author} {\bibfnamefont
  {Y.}~\bibnamefont {Takano}},\ }\bibfield  {title} {\bibinfo {title} {Phase
  diagram, magnetic properties and specific heat of
  {Mn}$_{1-x}${Fe}$_{x}${PS}$_{3}$},\ }\href
  {https://doi.org/10.1016/j.jallcom.2007.06.063} {\bibfield  {journal}
  {\bibinfo  {journal} {Journal of Alloys and Compounds}\ }\textbf {\bibinfo
  {volume} {460}},\ \bibinfo {pages} {668} (\bibinfo {year}
  {2008})}\BibitemShut {NoStop}%
\bibitem [{\citenamefont {Rao}\ and\ \citenamefont
  {Raychaudhuri}(1992)}]{RRao1992}%
  \BibitemOpen
  \bibfield  {author} {\bibinfo {author} {\bibfnamefont {R.~R.}\ \bibnamefont
  {Rao}}\ and\ \bibinfo {author} {\bibfnamefont {A.~K.}\ \bibnamefont
  {Raychaudhuri}},\ }\bibfield  {title} {\bibinfo {title} {Magnetic studies of
  a mixed antiferromagnetic system {Fe}$_{1-x}${Ni}$_{x}${PS}$_{3}$},\ }\href
  {https://doi.org/10.1016/0022-3697(92)90103-k} {\bibfield  {journal}
  {\bibinfo  {journal} {Journal of Physics and Chemistry of Solids}\ }\textbf
  {\bibinfo {volume} {53}},\ \bibinfo {pages} {577} (\bibinfo {year}
  {1992})}\BibitemShut {NoStop}%
\bibitem [{\citenamefont {Shemerliuk}\ \emph {et~al.}(2021)\citenamefont
  {Shemerliuk}, \citenamefont {Wolter}, \citenamefont {Cao}, \citenamefont
  {Zhou}, \citenamefont {Yang}, \citenamefont {Büchner},\ and\ \citenamefont
  {Aswartham}}]{Shemerliuk2021}%
  \BibitemOpen
  \bibfield  {author} {\bibinfo {author} {\bibfnamefont {Y.}~\bibnamefont
  {Shemerliuk}}, \bibinfo {author} {\bibfnamefont {A.~U.~B.}\ \bibnamefont
  {Wolter}}, \bibinfo {author} {\bibfnamefont {G.}~\bibnamefont {Cao}},
  \bibinfo {author} {\bibfnamefont {Y.}~\bibnamefont {Zhou}}, \bibinfo {author}
  {\bibfnamefont {Z.}~\bibnamefont {Yang}}, \bibinfo {author} {\bibfnamefont
  {B.}~\bibnamefont {Büchner}},\ and\ \bibinfo {author} {\bibfnamefont
  {S.}~\bibnamefont {Aswartham}},\ }\bibfield  {title} {\bibinfo {title}
  {Tuning magnetic and transport properties in
  ({Mn}$_{1-x}${Ni}$_{x}$)$_2${P}$_2${S}$_6$ single crystals},\ }\href@noop {}
  {\bibfield  {journal} {\bibinfo  {journal} {preprint}\ } (\bibinfo {year}
  {2021})}\BibitemShut {NoStop}%
\bibitem [{\citenamefont {Rule}\ \emph {et~al.}(2002)\citenamefont {Rule},
  \citenamefont {Kennedy}, \citenamefont {Goossens}, \citenamefont {Mulders},\
  and\ \citenamefont {Hicks}}]{Rule2002}%
  \BibitemOpen
  \bibfield  {author} {\bibinfo {author} {\bibfnamefont {K.}~\bibnamefont
  {Rule}}, \bibinfo {author} {\bibfnamefont {S.}~\bibnamefont {Kennedy}},
  \bibinfo {author} {\bibfnamefont {D.}~\bibnamefont {Goossens}}, \bibinfo
  {author} {\bibfnamefont {A.}~\bibnamefont {Mulders}},\ and\ \bibinfo {author}
  {\bibfnamefont {T.}~\bibnamefont {Hicks}},\ }\bibfield  {title} {\bibinfo
  {title} {Contrasting antiferromagnetic order between {FePS}$_{3}$ and
  {MnPS}$_{3}$},\ }\href {https://doi.org/10.1007/s003390201363} {\bibfield
  {journal} {\bibinfo  {journal} {Applied Physics A: Materials Science {\&}
  Processing}\ }\textbf {\bibinfo {volume} {74}},\ \bibinfo {pages} {s811}
  (\bibinfo {year} {2002})}\BibitemShut {NoStop}%
\bibitem [{\citenamefont {Taylor}\ \emph {et~al.}(1973)\citenamefont {Taylor},
  \citenamefont {Steger},\ and\ \citenamefont {Wold}}]{Taylor1973}%
  \BibitemOpen
  \bibfield  {author} {\bibinfo {author} {\bibfnamefont {B.~E.}\ \bibnamefont
  {Taylor}}, \bibinfo {author} {\bibfnamefont {J.}~\bibnamefont {Steger}},\
  and\ \bibinfo {author} {\bibfnamefont {A.}~\bibnamefont {Wold}},\ }\bibfield
  {title} {\bibinfo {title} {Preparation and properties of some transition
  metal phosphorus trisulfide compounds},\ }\href
  {https://doi.org/10.1016/0022-4596(73)90175-8} {\bibfield  {journal}
  {\bibinfo  {journal} {Journal of Solid State Chemistry}\ }\textbf {\bibinfo
  {volume} {7}},\ \bibinfo {pages} {461} (\bibinfo {year} {1973})}\BibitemShut
  {NoStop}%
\bibitem [{Note1()}]{Note1}%
  \BibitemOpen
  \bibinfo {note} {SAINT(V8.30A), Bruker AXS Inc., Madison, Wisconsin, USA
  (2017)}\BibitemShut {NoStop}%
\bibitem [{Note2()}]{Note2}%
  \BibitemOpen
  \bibinfo {note} {Bruker, APEX3 v2018.1-0, Bruker AXS Inc., Madison,
  Wisconsin, USA (2017)}\BibitemShut {NoStop}%
\bibitem [{\citenamefont {Krause}\ \emph {et~al.}(2015)\citenamefont {Krause},
  \citenamefont {Herbst-Irmer}, \citenamefont {Sheldrick},\ and\ \citenamefont
  {Stalke}}]{Krause2015}%
  \BibitemOpen
  \bibfield  {author} {\bibinfo {author} {\bibfnamefont {L.}~\bibnamefont
  {Krause}}, \bibinfo {author} {\bibfnamefont {R.}~\bibnamefont
  {Herbst-Irmer}}, \bibinfo {author} {\bibfnamefont {G.~M.}\ \bibnamefont
  {Sheldrick}},\ and\ \bibinfo {author} {\bibfnamefont {D.}~\bibnamefont
  {Stalke}},\ }\bibfield  {title} {\bibinfo {title} {Comparison of silver and
  molybdenum microfocus x-ray sources for single-crystal structure
  determination},\ }\href {https://doi.org/10.1107/s1600576714022985}
  {\bibfield  {journal} {\bibinfo  {journal} {Journal of Applied
  Crystallography}\ }\textbf {\bibinfo {volume} {48}},\ \bibinfo {pages} {3}
  (\bibinfo {year} {2015})}\BibitemShut {NoStop}%
\bibitem [{\citenamefont {Palatinus}\ and\ \citenamefont
  {Chapuis}(2007)}]{Palatinus2007}%
  \BibitemOpen
  \bibfield  {author} {\bibinfo {author} {\bibfnamefont {L.}~\bibnamefont
  {Palatinus}}\ and\ \bibinfo {author} {\bibfnamefont {G.}~\bibnamefont
  {Chapuis}},\ }\bibfield  {title} {\bibinfo {title} {{SUPERFLIP} {\textendash}
  a computer program for the solution of crystal structures by charge flipping
  in arbitrary dimensions},\ }\href {https://doi.org/10.1107/s0021889807029238}
  {\bibfield  {journal} {\bibinfo  {journal} {Journal of Applied
  Crystallography}\ }\textbf {\bibinfo {volume} {40}},\ \bibinfo {pages} {786}
  (\bibinfo {year} {2007})}\BibitemShut {NoStop}%
\bibitem [{\citenamefont {Pet{\v{r}}{\'{\i}}{\v{c}}ek}\ \emph
  {et~al.}(2014)\citenamefont {Pet{\v{r}}{\'{\i}}{\v{c}}ek}, \citenamefont
  {Du{\v{s}}ek},\ and\ \citenamefont {Palatinus}}]{Petricek2014}%
  \BibitemOpen
  \bibfield  {author} {\bibinfo {author} {\bibfnamefont {V.}~\bibnamefont
  {Pet{\v{r}}{\'{\i}}{\v{c}}ek}}, \bibinfo {author} {\bibfnamefont
  {M.}~\bibnamefont {Du{\v{s}}ek}},\ and\ \bibinfo {author} {\bibfnamefont
  {L.}~\bibnamefont {Palatinus}},\ }\bibfield  {title} {\bibinfo {title}
  {Crystallographic computing system {JANA2006}: {G}eneral features},\ }\href
  {https://doi.org/10.1515/zkri-2014-1737} {\bibfield  {journal} {\bibinfo
  {journal} {Zeitschrift für Kristallographie - Crystalline Materials}\
  }\textbf {\bibinfo {volume} {229}},\ \bibinfo {pages} {345} (\bibinfo {year}
  {2014})}\BibitemShut {NoStop}%
\bibitem [{\citenamefont {Sheldrick}(2007)}]{Sheldrick2007}%
  \BibitemOpen
  \bibfield  {author} {\bibinfo {author} {\bibfnamefont {G.~M.}\ \bibnamefont
  {Sheldrick}},\ }\bibfield  {title} {\bibinfo {title} {A short history of
  {SHELX}},\ }\href {https://doi.org/10.1107/s0108767307043930} {\bibfield
  {journal} {\bibinfo  {journal} {Acta Crystallographica Section A Foundations
  of Crystallography}\ }\textbf {\bibinfo {volume} {64}},\ \bibinfo {pages}
  {112} (\bibinfo {year} {2007})}\BibitemShut {NoStop}%
\bibitem [{\citenamefont {Klingen}\ \emph {et~al.}(1973)\citenamefont
  {Klingen}, \citenamefont {Eulenberger},\ and\ \citenamefont
  {Hahn}}]{Klingen1973}%
  \BibitemOpen
  \bibfield  {author} {\bibinfo {author} {\bibfnamefont {W.}~\bibnamefont
  {Klingen}}, \bibinfo {author} {\bibfnamefont {G.}~\bibnamefont
  {Eulenberger}},\ and\ \bibinfo {author} {\bibfnamefont {H.}~\bibnamefont
  {Hahn}},\ }\bibfield  {title} {\bibinfo {title} {\"{U}ber die
  {K}ristallstrukturen von {Fe}$_{2}${P}$_{2}${Se}$_{6}$ und
  {Fe}$_{2}${P}$_{2}${S}$_{6}$},\ }\href
  {https://doi.org/10.1002/zaac.19734010113} {\bibfield  {journal} {\bibinfo
  {journal} {Zeitschrift f\"{u}r anorganische und allgemeine Chemie}\ }\textbf
  {\bibinfo {volume} {401}},\ \bibinfo {pages} {97} (\bibinfo {year}
  {1973})}\BibitemShut {NoStop}%
\bibitem [{\citenamefont {Momma}\ and\ \citenamefont
  {Izumi}(2011)}]{KMomma2011}%
  \BibitemOpen
  \bibfield  {author} {\bibinfo {author} {\bibfnamefont {K.}~\bibnamefont
  {Momma}}\ and\ \bibinfo {author} {\bibfnamefont {F.}~\bibnamefont {Izumi}},\
  }\bibfield  {title} {\bibinfo {title} {{VESTA} 3 for three-dimensional
  visualization of crystal, volumetric and morphology data},\ }\href
  {https://doi.org/10.1107/s0021889811038970} {\bibfield  {journal} {\bibinfo
  {journal} {Journal of Applied Crystallography}\ }\textbf {\bibinfo {volume}
  {44}},\ \bibinfo {pages} {1272} (\bibinfo {year} {2011})}\BibitemShut
  {NoStop}%
\bibitem [{\citenamefont {Ouvrard}\ \emph {et~al.}(1985)\citenamefont
  {Ouvrard}, \citenamefont {Brec},\ and\ \citenamefont {Rouxel}}]{Ouvrard1985}%
  \BibitemOpen
  \bibfield  {author} {\bibinfo {author} {\bibfnamefont {G.}~\bibnamefont
  {Ouvrard}}, \bibinfo {author} {\bibfnamefont {R.}~\bibnamefont {Brec}},\ and\
  \bibinfo {author} {\bibfnamefont {J.}~\bibnamefont {Rouxel}},\ }\bibfield
  {title} {\bibinfo {title} {Structural determination of some {MPS}$_{3}$
  layered phases ({M} = {Mn}, {Fe}, {Co}, {Ni} and {Cd})},\ }\href
  {https://doi.org/10.1016/0025-5408(85)90092-3} {\bibfield  {journal}
  {\bibinfo  {journal} {Materials Research Bulletin}\ }\textbf {\bibinfo
  {volume} {20}},\ \bibinfo {pages} {1181} (\bibinfo {year}
  {1985})}\BibitemShut {NoStop}%
\bibitem [{\citenamefont {Dioguardi}\ \emph {et~al.}(2020)\citenamefont
  {Dioguardi}, \citenamefont {Selter}, \citenamefont {Peeck}, \citenamefont
  {Aswartham}, \citenamefont {Sturza}, \citenamefont {Murugesan}, \citenamefont
  {Eldeeb}, \citenamefont {Hozoi}, \citenamefont {Büchner},\ and\
  \citenamefont {Grafe}}]{Dioguardi2020}%
  \BibitemOpen
  \bibfield  {author} {\bibinfo {author} {\bibfnamefont {A.~P.}\ \bibnamefont
  {Dioguardi}}, \bibinfo {author} {\bibfnamefont {S.}~\bibnamefont {Selter}},
  \bibinfo {author} {\bibfnamefont {U.}~\bibnamefont {Peeck}}, \bibinfo
  {author} {\bibfnamefont {S.}~\bibnamefont {Aswartham}}, \bibinfo {author}
  {\bibfnamefont {M.-I.}\ \bibnamefont {Sturza}}, \bibinfo {author}
  {\bibfnamefont {R.}~\bibnamefont {Murugesan}}, \bibinfo {author}
  {\bibfnamefont {M.~S.}\ \bibnamefont {Eldeeb}}, \bibinfo {author}
  {\bibfnamefont {L.}~\bibnamefont {Hozoi}}, \bibinfo {author} {\bibfnamefont
  {B.}~\bibnamefont {Büchner}},\ and\ \bibinfo {author} {\bibfnamefont
  {H.-J.}\ \bibnamefont {Grafe}},\ }\bibfield  {title} {\bibinfo {title}
  {Quasi-two-dimensional magnetic correlations in {Ni}$_{2}${P}$_{2}${S}$_{6}$
  probed by $^{31}${P} {NMR}},\ }\href
  {https://doi.org/10.1103/physrevb.102.064429} {\bibfield  {journal} {\bibinfo
   {journal} {Physical Review B}\ }\textbf {\bibinfo {volume} {102}},\ \bibinfo
  {pages} {064429} (\bibinfo {year} {2020})}\BibitemShut {NoStop}%
\bibitem [{\citenamefont {Goossens}\ \emph {et~al.}(2011)\citenamefont
  {Goossens}, \citenamefont {James}, \citenamefont {Dong}, \citenamefont
  {Whitfield}, \citenamefont {Nor{\'{e}}n},\ and\ \citenamefont
  {Withers}}]{Goossens2011}%
  \BibitemOpen
  \bibfield  {author} {\bibinfo {author} {\bibfnamefont {D.~J.}\ \bibnamefont
  {Goossens}}, \bibinfo {author} {\bibfnamefont {D.}~\bibnamefont {James}},
  \bibinfo {author} {\bibfnamefont {J.}~\bibnamefont {Dong}}, \bibinfo {author}
  {\bibfnamefont {R.~E.}\ \bibnamefont {Whitfield}}, \bibinfo {author}
  {\bibfnamefont {L.}~\bibnamefont {Nor{\'{e}}n}},\ and\ \bibinfo {author}
  {\bibfnamefont {R.~L.}\ \bibnamefont {Withers}},\ }\bibfield  {title}
  {\bibinfo {title} {Local order in layered {NiPS}$_{3}$ and
  {Ni}$_{0.7}${Mg}$_{0.3}${PS}$_{3}$},\ }\href
  {https://doi.org/10.1088/0953-8984/23/6/065401} {\bibfield  {journal}
  {\bibinfo  {journal} {Journal of Physics: Condensed Matter}\ }\textbf
  {\bibinfo {volume} {23}},\ \bibinfo {pages} {065401} (\bibinfo {year}
  {2011})}\BibitemShut {NoStop}%
\bibitem [{\citenamefont {Shannon}(1976)}]{RShannon1976}%
  \BibitemOpen
  \bibfield  {author} {\bibinfo {author} {\bibfnamefont {R.~D.}\ \bibnamefont
  {Shannon}},\ }\bibfield  {title} {\bibinfo {title} {Revised effective ionic
  radii and systematic studies of interatomic distances in halides and
  chalcogenides},\ }\href {https://doi.org/10.1107/s0567739476001551}
  {\bibfield  {journal} {\bibinfo  {journal} {Acta Crystallographica Section
  A}\ }\textbf {\bibinfo {volume} {32}},\ \bibinfo {pages} {751} (\bibinfo
  {year} {1976})}\BibitemShut {NoStop}%
\bibitem [{\citenamefont {Vegard}(1921)}]{Vegard1921}%
  \BibitemOpen
  \bibfield  {author} {\bibinfo {author} {\bibfnamefont {L.}~\bibnamefont
  {Vegard}},\ }\bibfield  {title} {\bibinfo {title} {{D}ie {K}onstitution der
  {M}ischkristalle und die {R}aumf\"{u}llung der {A}tome},\ }\href
  {https://doi.org/10.1007/bf01349680} {\bibfield  {journal} {\bibinfo
  {journal} {Zeitschrift fuer Physik}\ }\textbf {\bibinfo {volume} {5}},\
  \bibinfo {pages} {17} (\bibinfo {year} {1921})}\BibitemShut {NoStop}%
\bibitem [{\citenamefont {Wildes}\ \emph {et~al.}(2015)\citenamefont {Wildes},
  \citenamefont {Simonet}, \citenamefont {Ressouche}, \citenamefont {McIntyre},
  \citenamefont {Avdeev}, \citenamefont {Suard}, \citenamefont {Kimber},
  \citenamefont {Lan{\c{c}}on}, \citenamefont {Pepe}, \citenamefont
  {Moubaraki},\ and\ \citenamefont {Hicks}}]{AWildes2015}%
  \BibitemOpen
  \bibfield  {author} {\bibinfo {author} {\bibfnamefont {A.~R.}\ \bibnamefont
  {Wildes}}, \bibinfo {author} {\bibfnamefont {V.}~\bibnamefont {Simonet}},
  \bibinfo {author} {\bibfnamefont {E.}~\bibnamefont {Ressouche}}, \bibinfo
  {author} {\bibfnamefont {G.~J.}\ \bibnamefont {McIntyre}}, \bibinfo {author}
  {\bibfnamefont {M.}~\bibnamefont {Avdeev}}, \bibinfo {author} {\bibfnamefont
  {E.}~\bibnamefont {Suard}}, \bibinfo {author} {\bibfnamefont {S.~A.~J.}\
  \bibnamefont {Kimber}}, \bibinfo {author} {\bibfnamefont {D.}~\bibnamefont
  {Lan{\c{c}}on}}, \bibinfo {author} {\bibfnamefont {G.}~\bibnamefont {Pepe}},
  \bibinfo {author} {\bibfnamefont {B.}~\bibnamefont {Moubaraki}},\ and\
  \bibinfo {author} {\bibfnamefont {T.~J.}\ \bibnamefont {Hicks}},\ }\bibfield
  {title} {\bibinfo {title} {Magnetic structure of the quasi-two-dimensional
  antiferromagnet {NiPS}$_{3}$},\ }\href
  {https://doi.org/10.1103/physrevb.92.224408} {\bibfield  {journal} {\bibinfo
  {journal} {Physical Review B}\ }\textbf {\bibinfo {volume} {92}},\ \bibinfo
  {pages} {224408} (\bibinfo {year} {2015})}\BibitemShut {NoStop}%
\bibitem [{\citenamefont {Berthier}\ \emph {et~al.}(1978)\citenamefont
  {Berthier}, \citenamefont {Chabre},\ and\ \citenamefont
  {Minier}}]{CBerthier1978}%
  \BibitemOpen
  \bibfield  {author} {\bibinfo {author} {\bibfnamefont {C.}~\bibnamefont
  {Berthier}}, \bibinfo {author} {\bibfnamefont {Y.}~\bibnamefont {Chabre}},\
  and\ \bibinfo {author} {\bibfnamefont {M.}~\bibnamefont {Minier}},\
  }\bibfield  {title} {\bibinfo {title} {Nmr investigation of the layered
  transition metal phosphorus trichalcogenides and the intercalation compounds
  {Li}$_{x}${NiPS}$_{3}$},\ }\href
  {https://doi.org/10.1016/0038-1098(78)90434-9} {\bibfield  {journal}
  {\bibinfo  {journal} {Solid State Communications}\ }\textbf {\bibinfo
  {volume} {28}},\ \bibinfo {pages} {327} (\bibinfo {year} {1978})}\BibitemShut
  {NoStop}%
\bibitem [{\citenamefont {Jernberg}\ \emph {et~al.}(1984)\citenamefont
  {Jernberg}, \citenamefont {Bjarman},\ and\ \citenamefont
  {Wäppling}}]{Jernberg1984}%
  \BibitemOpen
  \bibfield  {author} {\bibinfo {author} {\bibfnamefont {P.}~\bibnamefont
  {Jernberg}}, \bibinfo {author} {\bibfnamefont {S.}~\bibnamefont {Bjarman}},\
  and\ \bibinfo {author} {\bibfnamefont {R.}~\bibnamefont {Wäppling}},\
  }\bibfield  {title} {\bibinfo {title} {{FePS}$_{3}$: {A} first-order phase
  transition in a {\textquotedblleft}{2D}{\textquotedblright} ising
  antiferromagnet},\ }\href {https://doi.org/10.1016/0304-8853(84)90355-x}
  {\bibfield  {journal} {\bibinfo  {journal} {Journal of Magnetism and Magnetic
  Materials}\ }\textbf {\bibinfo {volume} {46}},\ \bibinfo {pages} {178}
  (\bibinfo {year} {1984})}\BibitemShut {NoStop}%
\bibitem [{\citenamefont {Kurosawa}\ \emph {et~al.}(1983)\citenamefont
  {Kurosawa}, \citenamefont {Saito},\ and\ \citenamefont
  {Yamaguchi}}]{KKurosawa1983}%
  \BibitemOpen
  \bibfield  {author} {\bibinfo {author} {\bibfnamefont {K.}~\bibnamefont
  {Kurosawa}}, \bibinfo {author} {\bibfnamefont {S.}~\bibnamefont {Saito}},\
  and\ \bibinfo {author} {\bibfnamefont {Y.}~\bibnamefont {Yamaguchi}},\
  }\bibfield  {title} {\bibinfo {title} {Neutron diffraction study on
  {MnPS}$_{3}$ and {FePS}$_{3}$},\ }\href
  {https://doi.org/10.1143/jpsj.52.3919} {\bibfield  {journal} {\bibinfo
  {journal} {Journal of the Physical Society of Japan}\ }\textbf {\bibinfo
  {volume} {52}},\ \bibinfo {pages} {3919} (\bibinfo {year}
  {1983})}\BibitemShut {NoStop}%
\bibitem [{\citenamefont {Rule}\ \emph {et~al.}(2007)\citenamefont {Rule},
  \citenamefont {McIntyre}, \citenamefont {Kennedy},\ and\ \citenamefont
  {Hicks}}]{KRule2007}%
  \BibitemOpen
  \bibfield  {author} {\bibinfo {author} {\bibfnamefont {K.~C.}\ \bibnamefont
  {Rule}}, \bibinfo {author} {\bibfnamefont {G.~J.}\ \bibnamefont {McIntyre}},
  \bibinfo {author} {\bibfnamefont {S.~J.}\ \bibnamefont {Kennedy}},\ and\
  \bibinfo {author} {\bibfnamefont {T.~J.}\ \bibnamefont {Hicks}},\ }\bibfield
  {title} {\bibinfo {title} {Single-crystal and powder neutron diffraction
  experiments on {FePS}$_{3}$: {S}earch for the magnetic structure},\ }\href
  {https://doi.org/10.1103/physrevb.76.134402} {\bibfield  {journal} {\bibinfo
  {journal} {Physical Review B}\ }\textbf {\bibinfo {volume} {76}},\ \bibinfo
  {pages} {134402} (\bibinfo {year} {2007})}\BibitemShut {NoStop}%
\bibitem [{\citenamefont {Blundell}(2001)}]{Blundell2001}%
  \BibitemOpen
  \bibfield  {author} {\bibinfo {author} {\bibfnamefont {S.}~\bibnamefont
  {Blundell}},\ }\href
  {https://www.ebook.de/de/product/3249307/stephen_department_of_physics_university_of_oxford_blundell_magnetism_in_condensed_matter.html}
  {\emph {\bibinfo {title} {Magnetism in Condensed Matter}}}\ (\bibinfo
  {publisher} {Oxford University Press},\ \bibinfo {year} {2001})\BibitemShut
  {NoStop}%
\bibitem [{\citenamefont {Akimitsu}\ \emph {et~al.}(1970)\citenamefont
  {Akimitsu}, \citenamefont {Ishikawa},\ and\ \citenamefont
  {Endoh}}]{JAkimitsu1970}%
  \BibitemOpen
  \bibfield  {author} {\bibinfo {author} {\bibfnamefont {J.}~\bibnamefont
  {Akimitsu}}, \bibinfo {author} {\bibfnamefont {Y.}~\bibnamefont {Ishikawa}},\
  and\ \bibinfo {author} {\bibfnamefont {Y.}~\bibnamefont {Endoh}},\ }\bibfield
   {title} {\bibinfo {title} {On the two-dimensional antiferromagnetic
  character of {MnTiO}$_{3}$},\ }\href
  {https://doi.org/10.1016/0038-1098(70)90578-8} {\bibfield  {journal}
  {\bibinfo  {journal} {Solid State Communications}\ }\textbf {\bibinfo
  {volume} {8}},\ \bibinfo {pages} {87} (\bibinfo {year} {1970})}\BibitemShut
  {NoStop}%
\bibitem [{\citenamefont {Birgeneau}\ \emph {et~al.}(1969)\citenamefont
  {Birgeneau}, \citenamefont {Guggenheim},\ and\ \citenamefont
  {Shirane}}]{RBirgeneau1969}%
  \BibitemOpen
  \bibfield  {author} {\bibinfo {author} {\bibfnamefont {R.~J.}\ \bibnamefont
  {Birgeneau}}, \bibinfo {author} {\bibfnamefont {H.~J.}\ \bibnamefont
  {Guggenheim}},\ and\ \bibinfo {author} {\bibfnamefont {G.}~\bibnamefont
  {Shirane}},\ }\bibfield  {title} {\bibinfo {title} {Neutron scattering from
  {K}$_{2}${NiF}$_{4}$: {A} two-dimensional heisenberg antiferromagnet},\
  }\href {https://doi.org/10.1103/physrevlett.22.720} {\bibfield  {journal}
  {\bibinfo  {journal} {Physical Review Letters}\ }\textbf {\bibinfo {volume}
  {22}},\ \bibinfo {pages} {720} (\bibinfo {year} {1969})}\BibitemShut
  {NoStop}%
\bibitem [{\citenamefont {Lueken}(1999)}]{Lueken1999}%
  \BibitemOpen
  \bibfield  {author} {\bibinfo {author} {\bibfnamefont {H.}~\bibnamefont
  {Lueken}},\ }\href {https://doi.org/10.1007/978-3-322-80118-0} {\emph
  {\bibinfo {title} {Magnetochemie}}}\ (\bibinfo  {publisher} {Vieweg+Teubner
  Verlag},\ \bibinfo {year} {1999})\BibitemShut {NoStop}%
\bibitem [{\citenamefont {Jenjeti}\ \emph {et~al.}(2018)\citenamefont
  {Jenjeti}, \citenamefont {Kumar}, \citenamefont {Austeria},\ and\
  \citenamefont {Sampath}}]{Jenjeti2018}%
  \BibitemOpen
  \bibfield  {author} {\bibinfo {author} {\bibfnamefont {R.~N.}\ \bibnamefont
  {Jenjeti}}, \bibinfo {author} {\bibfnamefont {R.}~\bibnamefont {Kumar}},
  \bibinfo {author} {\bibfnamefont {M.~P.}\ \bibnamefont {Austeria}},\ and\
  \bibinfo {author} {\bibfnamefont {S.}~\bibnamefont {Sampath}},\ }\bibfield
  {title} {\bibinfo {title} {Field effect transistor based on layered
  {NiPS}$_{3}$},\ }\href {https://doi.org/10.1038/s41598-018-26522-1}
  {\bibfield  {journal} {\bibinfo  {journal} {Scientific Reports}\ }\textbf
  {\bibinfo {volume} {8}},\ \bibinfo {pages} {8586} (\bibinfo {year}
  {2018})}\BibitemShut {NoStop}%
\bibitem [{\citenamefont {Kumar}\ \emph {et~al.}(2019)\citenamefont {Kumar},
  \citenamefont {Jenjeti}, \citenamefont {Austeria},\ and\ \citenamefont
  {Sampath}}]{Kumar2019}%
  \BibitemOpen
  \bibfield  {author} {\bibinfo {author} {\bibfnamefont {R.}~\bibnamefont
  {Kumar}}, \bibinfo {author} {\bibfnamefont {R.~N.}\ \bibnamefont {Jenjeti}},
  \bibinfo {author} {\bibfnamefont {M.~P.}\ \bibnamefont {Austeria}},\ and\
  \bibinfo {author} {\bibfnamefont {S.}~\bibnamefont {Sampath}},\ }\bibfield
  {title} {\bibinfo {title} {Bulk and few-layer {MnPS}$_{3}$: {A} new candidate
  for field effect transistors and {UV} photodetectors},\ }\href
  {https://doi.org/10.1039/c8tc05011b} {\bibfield  {journal} {\bibinfo
  {journal} {Journal of Materials Chemistry C}\ }\textbf {\bibinfo {volume}
  {7}},\ \bibinfo {pages} {324} (\bibinfo {year} {2019})}\BibitemShut {NoStop}%
\end{thebibliography}%

\end{document}